\documentclass[a4paper,pra,twocolumn,floatfix,superscriptaddress,showpacs]{revtex4}

\usepackage[utf8]{inputenc}
\usepackage{t1enc,hyperref,amsmath,amssymb,xspace,graphicx,esint}
\usepackage{tikz}

\renewcommand{\d}{\mathrm{d}}
\renewcommand{\vec}[1]{\boldsymbol{#1}}
\renewcommand{\Re}{\text{Re}\/}
\renewcommand{\Im}{\text{Im}\/}

\newcommand{\mB}{\ensuremath{m_{\phi}}\xspace}
\newcommand{\nF}{\ensuremath{n_{\psi}}\xspace}
\newcommand{\mF}{\ensuremath{m_{\psi}}\xspace}

\newcommand{\gBB}{\ensuremath{g_{\phi\phi}}\xspace}
\newcommand{\ginv}{\ensuremath{g_{\phi\psi}^{-1}}\xspace}
\newcommand{\ainv}{\ensuremath{a_{\phi\psi}^{-1}}\xspace}
\newcommand{\aBB}{\ensuremath{a_{\phi\phi}}\xspace}
\newcommand{\gBF}{\ensuremath{g_{\phi\psi}}\xspace}
\newcommand{\gBBtilde}{\ensuremath{\tilde{g}_{\phi\phi}}\xspace}
\newcommand{\gBFtilde}{\ensuremath{\tilde{g}_{\phi\psi}}\xspace}
\newcommand{\aBF}{\ensuremath{a_{\phi\psi}}\xspace}
\newcommand{\rhoBEC}{\ensuremath{\rho_0}\xspace}
\newcommand{\tmatrix}{\ensuremath{\Gamma_{\phi\psi}}\xspace}
\newcommand{\self}{\ensuremath{\Sigma_{\psi}}\xspace}
\newcommand{\GBbare}{\ensuremath{ {G_\phi^{(0)}}  }\xspace}
\newcommand{\Gnormal}{\ensuremath{ {G_{11}^{\phi}} }\xspace}
\newcommand{\Ganomalous}{\ensuremath{ {G_{21}^{\phi}} }\xspace}
\newcommand{\GFbare}{\ensuremath{ {G_\psi^{(0)}}  }\xspace}
\newcommand{\GFhalf}{\ensuremath{ {G_\psi^{(1)}}  }\xspace}
\newcommand{\GFfull}{\ensuremath{  {G_\psi} }\xspace}
\newcommand{\GM}{\ensuremath{  {G_\xi} }\xspace}
\newcommand{\GMbare}{\ensuremath{ {G_\xi^{(0)}}  }\xspace}
\newcommand{\muB}{\ensuremath{\mu_\phi}\xspace}
\newcommand{\muF}{\ensuremath{\mu_\psi}\xspace}

\newcommand{\Apol}{\ensuremath{A_{\text{pol}}}\xspace}
\newcommand{\Seff}{\ensuremath{S_{\text{eff}}}\xspace}
\newcommand{\xmax}{\ensuremath{x_{\text{max}}}}
\newcommand{\Omegamin}{\ensuremath{\Omega_{\text{min}}}}
\newcommand{\meff}{\ensuremath{m^*}\xspace}
\newcommand{\Eatt}{\ensuremath{E_{\text{att}}}\xspace}
\newcommand{\Erep}{\ensuremath{E_{\text{rep}}}\xspace}
\newcommand{\Edim}{\ensuremath{E_{\text{dim}}}\xspace}
\newcommand{\nsct}{NSCT\xspace}
\newcommand{\sct}{SCT\xspace}
\newcommand{\nuM}{\ensuremath{\nu_{\text{M}}}\xspace}

\DeclareMathOperator{\atanh}{\text{atanh}}

\begin{document}
\title{Field-theoretical study of the Bose polaron}
\author{Steffen Patrick Rath}
\affiliation{Technische Universit\"at M\"unchen, James-Franck-Stra\ss e, 85748
Garching, Germany}
\author{Richard Schmidt}
\affiliation{Technische Universit\"at M\"unchen, James-Franck-Stra\ss e, 85748
Garching, Germany}
\affiliation{Department of Physics, Harvard University, Cambridge, MA 02138, 
USA}
\affiliation{ITAMP, Harvard-Smithsonian Center for Astrophysics, Cambridge, MA
02138, USA}
\date{\today}
\begin{abstract}
  We study the properties of the Bose polaron, an impurity strongly interacting
  with a Bose--Einstein condensate, using a field-theoretic approach and make
  predictions for the spectral function and various quasiparticle properties
  that can be tested in experiment. We find that most of the spectral weight is
  contained in a coherent attractive and a metastable repulsive polaron branch.
  We show that the qualitative behavior of the Bose polaron is well described by
  a non-selfconsistent T-matrix approximation by comparing analytical results to
  numerical data obtained from a fully selfconsistent T-matrix approach. The
  latter takes into account an infinite number of bosons excited from the
  condensate.
\end{abstract}
\maketitle

\section{Introduction}
\label{sec:introduction}

Quantum impurity problems are among the paradigms of quantum many-body physics.
While they provide one of the  conceptually simplest ways to probe their host
medium \cite{ande1967}, they are also of interest in their own right. A classic
example is the Landau--Pekar polaron~\cite{land1933,land1948}, a single charge
moving inside a polarizable medium. While hardly deforming the medium itself,
the impurity tends to self-localize for sufficiently strong interactions.
Fr\"ohlich derived an effective Hamiltonian for this particular
system~\cite{froh1954} which since has found much application also for
microscopically different systems. Girardeau was the first to study a neutral
impurity immersed in  superfluid $^4$He~\cite{gira1961}.  With the advent of
cold atoms, which in recent years have proven a versatile testbed for the study
of quantum many-body systems~\cite{bloc2008}, new possibilities for the study of
quantum impurities have emerged. The host may consist of bosons or fermions,
interactions are tunable via Feshbach resonances and even the dimensionality of
the system can be engineered~\cite{korr2011}.  The case of the Fermi polaron
where the host medium is an ideal Fermi gas has received considerable
experimental attention. In particular, the technique of inverse radiofrequency
(rf) spectroscopy has permitted the study of the Fermi polaron's spectral
function  both in two and three dimensions~\cite{schm2011,kohs2012,korr2011}. 

The Bose polaron problem, where an impurity interacts with a Bose--Einstein
condensate (BEC), has so far not been studied very extensively in experiment,
with the existing literature mostly focusing on weakly interacting systems
\cite{chik2000,cata2012,spet2012,scel2013}.  With the recent progress on the
identification and characterization of Feshbach resonances in ultracold
Bose-Fermi mixtures~\cite{ospe2006,wu2011,park2012,park2012a}, experiments with
impurities featuring widely tunable interactions
with a BEC now seem within reach. In step with experimental technology, most of
the theoretical literature on the Bose polaron has focused on weak interspecies
interactions. Two particularly well studied approaches are on the one hand
coupled mean field equations which permit the characterization of polaron
self-localization~\cite{astr2004,cucc2006,sach2006,kala2006,brud2008,blin2013},
and on the other hand the study of an effective Fr\"ohlich-type
Hamiltonian~\cite{temp2009,cast2011,cast2011a,cast2011b,cast2012,dase2013} which
has also been applied to the description of Bose--Fermi
mixtures~\cite{stoo1996,bijl2000,heis2000,mate2003,wang2006,enss2009}.

In contrast to these effective approaches which rely on the microscopic
repulsion between the impurity and the host atoms, we start from a microscopic
description of the impurity interacting with a homogeneous BEC via a Feshbach
resonance which takes into account the attractiveness of the
interactions.  Analyzing the excitation spectrum, we find two branches. While
the first is a stable, long-lived quasiparticle representing an attractive
polaron, the second can be interpreted as a repulsive polaron.  The finite
lifetime of the latter is a consequence of the attractive
interactions and it inhibits the preparation of the system in a stable strongly
repulsive state, thus complicating a quantum simulation of the Landau--Pekar
polaron paradigm.  To circumvent the possibility of an instability towards a
ground state with an inhomogeneous BEC, we consider a setup where the impurity
is driven from a weakly interacting initial state to a final state strongly
interacting with a homogeneous condensate. We provide
predictions for the momentum-resolved spectral function which can be
experimentally tested with available
technology~\cite{stew2008,schi2009,feld2011,kohs2012,korr2011}.  Within a
T-matrix approach, we find that to a surprisingly good approximation, the
selfenergy of the Bose polaron is proportional to the bare interspecies
T-matrix. We show that the leading correction arises not due to weak
boson--boson interactions, but due to a selfconsistent treatment of the
interaction of the impurity with the BEC which takes into account multiple
virtual excitations of bosons out of the condensate.

\section{The model}
\label{sec:model}

We consider a mobile impurity of mass $\mF$ which is immersed in a
Bose--Einstein condensate (BEC) of atoms of mass $\mB$ and density $n$ at
temperature $T=0$.
Although in the case of a single impurity its statistics is of no consequence,
we will nonetheless refer to it as a fermion, having in mind the Bose polaron
problem as a Bose-Fermi mixture in the limit of extreme population imbalance
$\nF/n\ll 1$~\footnote{Note that the presence of a \textit{finite} density of
fermions can induce a collapse or phase separation of the BEC beyond a critical
interaction strength. This instability is due to fermionic particle--hole
fluctuations giving rise to a Lindhard type softening of the Bogoliubov modes
\cite{wang2006,enss2009}. In our case, $\nF\to 0 $, and such an effect, which is
different from the self-localization aspect to be discussed below, does not take
place.}. We assume that the bosons interact weakly among each other while they
interact with the impurity via a, potentially strong, short-range potential as
appropriate for ultracold atoms close to a magnetically tunable Feshbach
resonance \cite{chin2010}.  In the case of an open-channel dominated Feshbach
resonance (a description appropriate for Feshbach resonances of arbitrary width
is discussed in Appendix~\ref{app:molecule_propagator}), the system can be
described using a contact interaction~\cite{bloc2008} and the corresponding
microscopic, euclidean action reads (throughout the paper, we use units in which
$\hbar=1$)
\begin{multline}
  S=\int\d\tau\int\d^3x\,\Bigg\{ 
  \varphi^*_{\vec{x},\tau}\left( \partial_\tau-\frac{1}{2\mB}\nabla^2-\mu_\phi \right)
  \varphi_{\vec{x},\tau}
   \\ 
+\psi^*_{\vec{x},\tau}\left( \partial_\tau-\frac{1}{2\mF}\nabla^2-\mu_\psi \right)
  \psi_{\vec{x},\tau}
   \\ 
  +\frac{\gBB}{2}|\varphi_{\vec{x},\tau}|^4
  +\gBFtilde|\varphi_{\vec{x},\tau}|^2|\psi_{\vec{x},\tau}|^2
  \Bigg\}\ ,
  \label{eq:microscopic_action}
\end{multline}
where $\varphi$ denotes the bosonic, $\psi$ the impurity field, and $\mu_\phi$
and $\mu_\psi$ the corresponding chemical potentials. The microscopic couplings
$\gBB$ and $\gBFtilde$ are related to the experimentally relevant scattering
lengths $\aBB$ and $\aBF$
via the solution of the two-body Lippmann--Schwinger equation. While the
resulting relation between $\gBB$ and $\aBB$ is trivial due to the weakness of
the boson--boson coupling and gives rise to the common identity $\gBB\approx
4\pi\aBB/\mB>0$, it is more complicated in the case of the interspecies coupling
where the scattering length may become arbitrarily large
(cf.~Appendix~\ref{app:regbose}).  In Eq.~\eqref{eq:microscopic_action} we allow
for an arbitrary mass imbalance $\alpha=\mF/\mB$. While for our numerical
results we assume the mass-balanced case, we show all analytical results for the
general case where $\alpha\neq 1$.

In the following, we use an effective action which is based on the
Bogoliubov approximation and may be obtained as follows. In a first step, we
exploit the shift symmetry of the path integral and make the usual replacement
$\varphi(\vec x, t)\mapsto\sqrt{\rhoBEC(\vec x)}+ \phi(\vec x, t)$ where
$\rhoBEC(\vec x)>0$ is the condensate density and $\phi(\vec{x},t)$ represents
the fluctuations around the condensate. We expand around a
homogeneous condensate and make the simplifying assumption
that the condensate density is not altered by the presence of the impurity,
i.e.,  we replace the condensate by its mean field value $\rhoBEC(\vec x)\equiv
\rhoBEC=n$. Following Bogoliubov's prescription, we discard all powers of the
bosonic fields higher than two. After some rearrangements, the resulting
effective action can be written as
\begin{widetext}
  \begin{multline}
    \Seff = 
\int_{p}\left\{ 
  \frac{1}{2}
  \begin{pmatrix}
    \phi^*_p \\ \phi_{-p}
  \end{pmatrix}
  \begin{pmatrix}
    -\left[\GBbare(-p)\right]^{-1} & \gBB\rhoBEC \\
    \gBB\rhoBEC & -\left[\GBbare(p)\right]^{-1}
  \end{pmatrix}
  \begin{pmatrix}
    \phi_p \\ \phi^*_{-p}
  \end{pmatrix}
  +\psi^*_p \left(-i\omega +\frac{\vec{p}^2}{2\mF}-\mu_\psi \right)\psi_p
  \right\}
  \\
  +\gBFtilde\int_x|\psi_x|^2 \left[ |\phi_x|^2+(\phi_x+\phi^*_x)\sqrt{\rhoBEC}
  +\rhoBEC \right] 
    \label{eq:Seff}
  \end{multline}
\end{widetext}
where we have introduced the shorthand notation $p=(\omega,\vec{p})$ and
$x=(\tau,\vec{x})$.
$\left[\GBbare(p)\right]^{-1}=i\omega-\vec{p}^2/2\mB+\gBB\rhoBEC$ is the inverse
bare bosonic propagator. The chemical potential $\muB=\gBB\rho_0$ follows
from the Hugenholtz--Pines relation~\cite{huge1959} evaluated at the mean field
level and vanishes in the case of a non-interacting BEC. The bosonic propagator
is obtained by inverting the matrix appearing in Eq.~\eqref{eq:Seff}. Only two
of the four matrix elements are independent, namely the normal and the anomalous
propagator which are given by~\cite{abri1975}
\begin{align}
  \Gnormal(i\omega,\vec{p})&=-\frac{i\omega+\frac{\vec{p}^2}{2\mB}+\gBB\rhoBEC}
  {\omega^2+\varepsilon_{\vec{p}}^2}
  \stackrel{\gBB\rightarrow 0}{\longrightarrow}
  \GBbare(i\omega,\vec{p})
  \label{eq:boson_propagator_a}
  \\
\Ganomalous(i\omega,\vec{p})&=\frac{\gBB\rhoBEC}
{\omega^2+\varepsilon_{\vec{p}}^2}
  \stackrel{\gBB\rightarrow 0}{\longrightarrow}
  0
  \label{eq:boson_propagator_b}
\end{align}
with the Bogoliubov dispersion
\begin{equation}
  \varepsilon_{\vec{p}}=\sqrt{\frac{\vec{p}^2}{2\mB}\left(
  \frac{\vec{p}^2}{2\mB}+2\gBB\rhoBEC \right)} \stackrel{\gBB\rightarrow
  0}{\longrightarrow}\frac{\vec p^2}{2\mB}\ .
  \label{eq:Bogoliubov}
\end{equation}
In Eqs.~\eqref{eq:boson_propagator_a}-\eqref{eq:Bogoliubov} we indicate the
limit of the non-interacting Bose gas which we will study below as a
limiting case of our model. The bare impurity propagator in Eq.~\eqref{eq:Seff}
has the usual form 
\begin{equation}
  \GFbare(i\omega,\vec{p})=\frac{1}{i\omega-\frac{\vec{p}^2}{2\mF}+\muF}\ .
  \label{eq:GF} 
\end{equation}

By using the effective action \eqref{eq:Seff} with a homogeneous BEC we neglect
that for strong enough interspecies interactions, the actual mean field solution
corresponding to the action~\eqref{eq:microscopic_action} may feature an
inhomogeneous condensate~\footnote{In contrast to the Fermi polaron where the
host medium is given by a non-interacting Fermi sea with low compressibility due
to Fermi pressure,  such a response of a BEC to the presence of a single
impurity may be significant since here the host medium is a highly compressible
fluid.}. In fact, self-localization of the impurity and deformations of the BEC
have been predicted for both microscopically repulsive and attractive
interactions~\cite{kupe1961,cucc2006,kala2006,brud2008,blin2013} as well as in
the context of a two-fluid Bose--Bose mixture \cite{sart2013}.  Our neglecting
these effects in the study of the model \eqref{eq:Seff} can be justified by
considering the involved time scales.  The deformation of the BEC requires the
displacement of mass which in weakly interacting BECs typically happens on a
time scale of $\tau_\text{BEC}\sim\hbar/(\gBB n)$~\cite{blin2013}. In this
article we are, however, interested in the excitation spectrum of the Bose
polaron which can be experimentally probed via inverse radio frequency
spectroscopy. Here the impurity atom is driven from a weakly to a strongly
interacting state. In such experiments the time scale is set by a small fraction
of the inverse Rabi frequency, which can be significantly smaller than
$\tau_\text{BEC}$ \cite{kohs2012}. In consequence, the rf response of the system
is determined by the spectral function of an impurity interacting with a
homogeneous condensate and our model is expected to apply even in the strongly
interacting regime. 

\section{Non-selfconsistent T-matrix approximation}
\label{sec:nsct}
\label{sec:tmatrix}

\begin{figure}[htbp]
  \centering
  \includegraphics{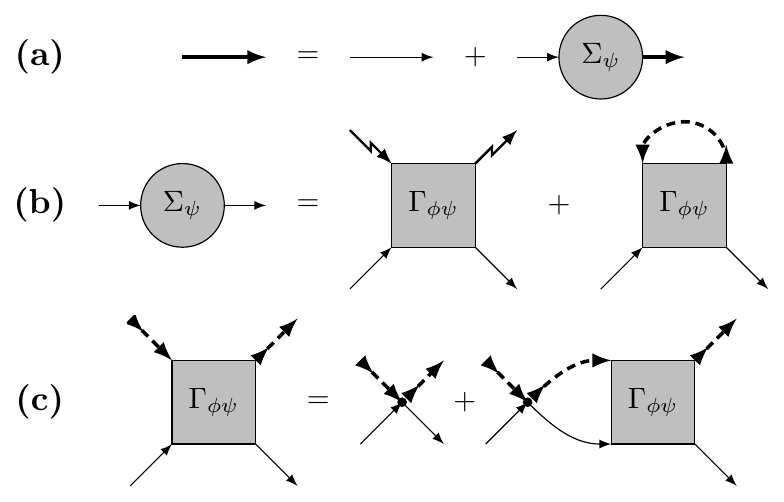}
  \caption{Diagrams corresponding to the non-selfconsistent T-matrix
  approximation: (a) Dyson's equation for the impurity Green's function. (b)
  Impurity selfenergy. (c) In-medium T-matrix equation. Solid thin lines
  stand for impurity propagators while dashed lines represent normal Bose
  propagators. The zig-zag lines denote in- and outgoing condensate particles.
  In the case of the selfconsistent T-matrix approximation discussed in
  Section~\ref{sec:SCT}, the solid thin line (bare fermion propagator) in the
  particle--particle loop in (c) is replaced by the thick line from (a)
  representing the dressed fermion propagator. }
  \label{fig:impurity_diagrams}
\end{figure}

The derivation of the \nsct equations is based on the perturbative expansion of
the impurity Green's function in the bare coupling constant \gBFtilde. Using
Dyson's equation~[cf.~Fig.~\ref{fig:impurity_diagrams}~(a)], one automatically
resums all one-particle reducible Feynman diagrams into the selfenergy \self.
Taking into account the diagrams shown in Fig.~\ref{fig:impurity_diagrams}~(b),
the latter reads 
\begin{multline}
  \self(\Omega,\vec{p}) = \rhoBEC\tmatrix(\Omega,\vec{p})
  \\
  -\left.\int_{\vec{k}}\int_\nu
  \Gnormal[i(\nu-\omega),\vec{k}-\vec{p}]\tmatrix(i\nu,\vec{k})\right|_{i\omega\mapsto \Omega+i0^+}
  \label{eq:def_selfenergy}
\end{multline}
where $\int_{\vec{k}}=\int\d^3k/(2\pi)^3$, $\int_\nu=\int\d\nu/2\pi$ and
\tmatrix is the in-medium T-matrix.
The first term corresponds to the Belyaev-type diagram in Fig.~\ref{fig:impurity_diagrams}~(b)~(\cite{bely1958},
cf.~\cite{shi1998,capo2008} for recent work on dilute Bose gases building on
Belyaev's technique) and represents a
mean field like interaction of the impurity with the condensate. In the case of
a Bose gas in the normal phase, this term is absent.   The second term reflects
the dressing of the impurity with bosons which are depleted from the condensate
due to the boson--boson interactions and vanishes in the case of a
noninteracting Bose gas. In a calculation at $T\neq0$ this term would equally
account for scattering off thermally depleted particles. Within the \nsct
approximation, the selfenergy only contains the class of two-particle reducible
ladder-type diagrams shown in Fig.~\ref{fig:impurity_diagrams}~(c). The T-matrix
\tmatrix is given by the infinite sum over this class of diagrams which
represents the repeated boson--impurity scattering in the $s$-wave
particle--particle channel,
\begin{equation}
  \tmatrix(\Omega,\vec{p})=\left[\ginv + L(\Omega,\vec{p}) \right]^{-1}
  \label{eq:tmatrix_eqn}
\end{equation}
with the pair propagator 
\begin{multline}
  L(\Omega,\vec{p}) = \int_{\vec{k}}\left\{ 
  \int_{\nu}\Gnormal( i \nu,\vec{k})\GFbare[i(\omega-\nu),\vec{p}-\vec{k}]
  \right. \\ \left.\left.
  -\frac{2\alpha \mB}{(\alpha+1)\vec{k}^2}
  \right\}\right|_{i\omega\mapsto \Omega+i0^+}\ .
  \label{eq:pair_propgator}
\end{multline}
We note that in the integral both free particle and Bogoliubov dispersions
appear. This reflects the possible scattering of the impurity off phonons with
linear dispersion at low scattering energies while at large energies it
effectively scatters off the ``fundamental'' bosons.  While the imaginary
frequencies used so far provide a particularly clear way of evaluating frequency
integrals, we ultimately want to compare with experimentally observable
dynamical properties such as rf spectra and quasiparticle properties. To
this end, we will consider all quantities at real frequencies obtained by
analytic continuation $i\omega\mapsto\omega + i0^+$ which yields the retarded
Green's function. As a convenient shorthand notation we introduce $\Omega =
\omega+\muF$ as already used in Eqs.~\eqref{eq:def_selfenergy} and
\eqref{eq:pair_propgator}.

The term in the second line of Eq.~\eqref{eq:pair_propgator} originates from the
renormalization of the bare coupling \gBFtilde. As discussed in detail
in Appendix~\ref{app:regbose}, the particle--particle loop has an ultraviolet
divergence which is regularized at a momentum cutoff scale $\Lambda$. The
dependence on $\Lambda$ is subsequently absorbed in the definition of the bare
coupling constant $ \tilde{g}_{\phi\psi}$, yielding the renormalized interspecies coupling
constant
\begin{equation}
  \ginv = \tilde{g}_{\phi\psi}^{-1} 
  + \int\frac{\d^3k}{(2\pi)^3}\frac{2\alpha \mB}{(\alpha+1)\vec{k}^2}
  =\frac{\alpha \mB}{2\pi(\alpha+1)}\aBF^{-1}\ ,
  \label{eq:gBF_renormalized}
\end{equation}
where $\alpha \mB/(\alpha+1)$ is the reduced mass for boson--impurity
collisions. 

The \nsct approximation has first been used to describe rf experiments on
imbalanced Fermi gases and the Fermi polaron (e.g.,
\cite{punk2007,comb2007,pera2008,mass2008}).  In the case of two-component
fermions it is equivalent to a leading order $1/N$ expansion \cite{enss2012} and
the Nozi{\`e}res--Schmitt-Rink approach \cite{nozi1985}.  In the context of the
Fermi polaron problem (where it turns out to be equivalent to a calculation
using a variational wave function~\cite{chev2006a}), the \nsct approximation has
successfully been applied~\cite{comb2007} and yields results for various
quasiparticle properties in quantitative agreement with experiments
\cite{schi2009,kohs2012} and state of the art diagrammatic Monte-Carlo methods
\cite{prok2008,prok2008b}.  Finally we mention that the \nsct approximation has
been employed by Fratini and coworkers \cite{frat2010} for the study of the
phase diagram of the Bose--Fermi action \eqref{eq:microscopic_action} in the
case of a finite boson and fermion density.  In this study the bosons were,
however, in the non-condensed phase.  Note that the ladder diagrams from
Fig.~\ref{fig:impurity_diagrams}~(c) incorporate the dominant pairing
fluctuations present in the system.  Their resummation is essential to obtain
the attractive and repulsive polaron features mentioned in the introduction.
These pairing fluctuations are taken into account neither in a diagrammatic
expansion based on the Fr{\"o}hlich
Hamiltonian~\cite{stoo1996,heis2000,mate2003,wang2006,enss2009,temp2009} nor in
mean field
approaches~\cite{astr2004,cucc2006,sach2006,kala2006,brud2008,blin2013}. 

In the case of the Bose polaron, there are two particularities compared to the
case of two-component fermions or a Bose-Fermi mixture in the normal
phase. First, the dashed lines appearing in the loop expressions in
Fig.~\ref{fig:impurity_diagrams} represent not simple bare boson propagators
but the mean field Green's functions including the condensate and featuring a
double-pole structure with Bogoliubov dispersion,
cf.~Eqs.~\eqref{eq:boson_propagator_a} and \eqref{eq:Bogoliubov}. Second, the
presence of the condensate leads to the appearance of additional diagrams where
the in- and outgoing lines of the T-matrix represent atoms  ejected out of or
injected into the condensate~\footnote{This results in a set of four different
T-matrices  which differ in the kind of incoming and outgoing bosonic particle
(condensate or non-condensate). Two of these T-matrices appear for instance in
the impurity selfenergy~[cf.~Fig.~\ref{fig:impurity_diagrams}~(b)]. It turns out
that within both the selfconsistent and the non-selfconsistent T-matrix
approximation all four vertices are represented by the same (amputated) vertex
$\tmatrix$.}.  Note that within the T-matrix approximation, the term $\gBFtilde
\rho_0$---as also implied by the notation for the effective action
\eqref{eq:Seff}---is included in the interaction part of the action instead of
being incorporated in the bare Fermi propagator or absorbed in the chemical
potential. These latter choices would lead to inconsistent expressions involving
both renormalized and unrenormalized couplings.  

Of course, the T-matrix approach is a non-perturbative approximation. In this
respect, beyond the second term in the self-energy in
Fig.-\ref{fig:impurity_diagrams}(b), there are more processes which contribute
and which we do not take into account. In fact, the diagrammatic structure of
our T-matrix approximation is
basically the same as in the description of a BEC within a T-matrix
approximation with respect to the boson--boson coupling (cf.~\cite{shi1998} and
references therein). For instance, like in the purely bosonic case, we ignore
diagrams in the selfenergy featuring anomalous Bose propagators or vertex
corrections. These processes cannot be included in the T-matrix approximation
scheme without causing inconsistencies such as double counting. The T-matrix
approach however correctly includes the pairing correlations responsible for the
leading instability in the case of ultracold atoms close to a Feshbach
resonance. Thus, it allows to access the strong-coupling regime and to reveal
the intricate physics beyond the Fröhlich polaron paradigm. Note that also in
the case of a weak boson--boson interaction the pairing channel remains
dominant. Processes featuring, for instance, anomalous propagators represent
only corrections. This is particularly apparent as the anomalous propagator
scales with $\gBB$ while the pairing instability survives the limit $\gBB\to0 $.
We thus expect the T-matrix approximation to be a valid description of the Bose
polaron in the regime where the host is a weakly interacting Bose gas.

Remarkably, the pair propagator $L(\Omega,\vec{p})$ can be determined fully
analytically for arbitrary frequencies, momenta, mass ratios and interaction
constants. This is an important feature also for the numerical calculation of
spectral functions as it reduces the numerical effort by orders of magnitude and
allows for a direct analytical continuation of the results to real frequencies.
Since the derivation is somewhat lengthy, we refer to Appendix \ref{app:tmatrix}
for the detailed calculation. However, to illustrate the general structure of
the solution, we show here the particularly simple result one obtains in the
special case $\alpha=1$ and $p=0$ where the real and imaginary part of the pair
propagator read 
\begin{equation} 
  \Re L(\Omega)=\frac{-1}{2\pi^2} \begin{cases}
    \sqrt{\gBB}-\frac{\arctan\sqrt{-1-\frac{\Omega}{\gBB}}}{\sqrt{-\gBB-\Omega}}\Omega & \Omega < -1
    \\[.7em] 
    \sqrt{\gBB}-\frac{\atanh\sqrt{1+\frac{\Omega}{\gBB}}}{\sqrt{\gBB +\Omega}}\Omega &
    -1\leq\Omega\leq 0 \\[.7em]
    \sqrt{\gBB}-\frac{\atanh\frac{1}{\sqrt{1+\frac{\Omega}{\gBB}}}}{\sqrt{\gBB +\Omega}}\Omega & 0 < \Omega
  \end{cases}
  \label{eq:ReL_p0_rho_units} 
\end{equation}
and
\begin{equation}
  \Im L(\Omega,0) = \frac{1}{4\pi}\frac{\Omega}{\sqrt{\gBB
  +\Omega}}\theta(\Omega) \ .
  \label{eq:ImL_alpha_1_p0_rho_units}
\end{equation}
In these equations as well as the remainder of this article, we adopt
units in which $\mB=n=1$, i.e., the length scale is set by the
mean interparticle spacing $\sim n^{-1/3}$ of the bosons. Occasionally
we will retain the explicit $m_\phi$ and $n$ (or $\rho_0$) dependence for
clarity. In Fig.~\ref{fig:L} in Appendix~\ref{app:tmatrix}, we plot the pair
propagator as a function of frequency and momentum (note that for calculational
convenience this Appendix uses a different set of units). In the limit
$\gBB\rightarrow0$ the pair propagator---now again for arbitrary mass
ratio---reduces to 
\begin{equation} L(\Omega,\vec{p}) = \frac{i}{4\pi}\left(
  \frac{2\alpha}{\alpha+1} \right)^{3/2}
  \sqrt{\Omega-\frac{\vec{p}^2}{2(\alpha+1)}+i0^+}\ .  \label{eq:Lvacuum}
\end{equation} 
The corresponding T-matrix obtained via Eq.~\eqref{eq:tmatrix_eqn} is identical
to the T-matrix in vacuum up to the appearance of the chemical potential
$\mu_\psi$ inside $\Omega$.

Having obtained the T-matrix $\tmatrix$ explicitly, we next turn to the
calculation of the corresponding impurity selfenergy. 
The frequency integral in Eq.~\eqref{eq:def_selfenergy} can be performed
analytically since the vanishing impurity density dictates that 
the T-matrix has all its singularities in the lower half plane when one uses
imaginary frequencies. The integration contour may thus be closed in the upper
half plane and the integral picks up a contribution from one of the boson
propagator's poles.  After analytic continuation, the
selfenergy~\eqref{eq:def_selfenergy} then reads
\begin{multline}
  \self(\Omega,\vec{p}) = \tmatrix(\Omega,\vec{p})
  \\
  -\int\frac{\d^3k}{(2\pi)^3}\frac{\sqrt{k^2(k^2+4\gBB)}
     -(k^2+2\gBB)}
     {2\sqrt{k^2(k^2+4\gBB)}}
     \\
     \times\tmatrix
     \left(\Omega-\frac{1}{2}\sqrt{k^2(k^2+4\gBB)},\vec{k}+\vec{p}\right)
       \ .
  \label{eq:selfenergy_integrated}
\end{multline}
The remaining momentum integral is readily evaluated numerically.

The main focus of this article is on the excitation spectrum of the impurity.
The key theoretical quantity to extract this spectrum is the spectral function
of the polaron given by
\begin{align}
  \Apol(\Omega,\vec{p}) &= -2\,\Im G_\psi^{\text{R}}(\Omega,\vec{p})
  \label{eq:spectral_fcts}
\end{align}
with the full retarded impurity Green's function
\begin{equation}
  G_\psi^{\text{R}}(\Omega,\vec{p}) = \frac{1}{\Omega -
  \frac{\vec{p}^2}{2\mF}-\self(\Omega,\vec{p})+i0^+}
  \label{eq:GFfull}
\end{equation}
as obtained from Dyson's equation depicted in
Fig.~\ref{fig:impurity_diagrams}~(a). We will suppress the superscript R in the
following and the retardedness of propagators will be implied by the use of the
real frequency argument $\Omega\equiv\omega+\mu_\psi$.Using our conventions, the
spectral function fulfills the sum rule
$(2\pi)^{-1}\int\d\Omega\,\Apol(\Omega,\vec{p})=1$ where the integration extends
over all frequencies.  

The excitation spectrum contained in the impurity spectral function
Eq.~\eqref{eq:spectral_fcts} is determined by the analytical structure of the
Green's function Eq.~\eqref{eq:GFfull} in the complex frequency plane. While
branch cuts correspond to an incoherent continuum of excitations, poles are
linked to the existence of well-defined quasiparticles that can be
characterized by a small set of key quantities \cite{abri1975}: (i) The
quasiparticle dispersion relation $E(p)$ is defined as the solution of
\begin{equation}
  E(p)-\frac{p^2}{2\mF}-\Re\self[E(p),p] = 0\ ,
  \label{eq:defOmega0}
\end{equation}
where in the isotropic case considered here quantities depend only on the
magnitude of the momentum $p=|\vec p|$ (From here on the symbol $p$
denotes the magnitude of momentum and not the four-momentum). (ii) The
(momentum dependent) spectral weight is given by
\begin{equation}
  Z(p)=\frac{1}{1-\partial_\Omega\Re\self[\Omega,p]}\bigg|_{\Omega=E(p)} \ .
  \label{eq:weight}
\end{equation}
(iii) The decay width is obtained from
\begin{equation}
  \gamma(p) = -Z(p)\,\Im\self[E(p),p]
  \label{eq:decaywidth}
\end{equation}
and (iv) the momentum dependent effective mass reads
\begin{equation}
  \meff(p) = \frac{p}{\partial_pE(p)}
  =\frac{1/Z(p)}{\frac{1}{\alpha}+\frac{1}{p}\left.\partial_p
  \Re\self[\Omega,p]\right|_{E(p)}}\
  .
  \label{eq:effmass}
\end{equation}
Eqs.~\eqref{eq:defOmega0}--\eqref{eq:effmass} provide an accurate description of
the quasiparticle properties as long as the poles of the Green's function
$G_\psi(\Omega,p)$ in the complex frequency plane are close to the real
axis. If this condition is violated, the interpretation of the poles as
well-defined quasiparticles starts to break down and the preciseness of
Eqs.~\eqref{eq:defOmega0}--\eqref{eq:effmass} depends on how well one satisfies
the assumption that $\Im\self$ remains constant across the width
of the quasiparticle peak as well as the smallness of the decay width compared to the quasiparticle energy. 

We now turn to the analysis of the impurity's spectral function $\Apol$ given by
Eqs.~\eqref{eq:spectral_fcts}, \eqref{eq:GFfull}, and
\eqref{eq:selfenergy_integrated}. When choosing $\alpha=1$, as done for all
plots in this article, we are left with $\Apol$ as a function of four
quantities: $\Omega$, $p$, $\aBF$ and \aBB. We choose $n^{1/3}\aBB=0.1$ for the
following plots which is actually about one order of magnitude stronger than
what is typical for weakly interacting Bose gases. While such strong
interactions have actually been reached in
experiment~\cite{webe2003,krae2006,navo2011,mako2013} and do themselves lead to
interesting physics beyond the scope of this article, our motivation for this
large value is quite mundane: it turns out that in the T-matrix approximation
the dependence of the spectral properties on \gBB is very weak. So, to make this
dependence clearly visible in our plots,  we choose this somewhat exaggerated
value. 
\begin{figure}[htbp]
  \centering
    \includegraphics[width=\linewidth]{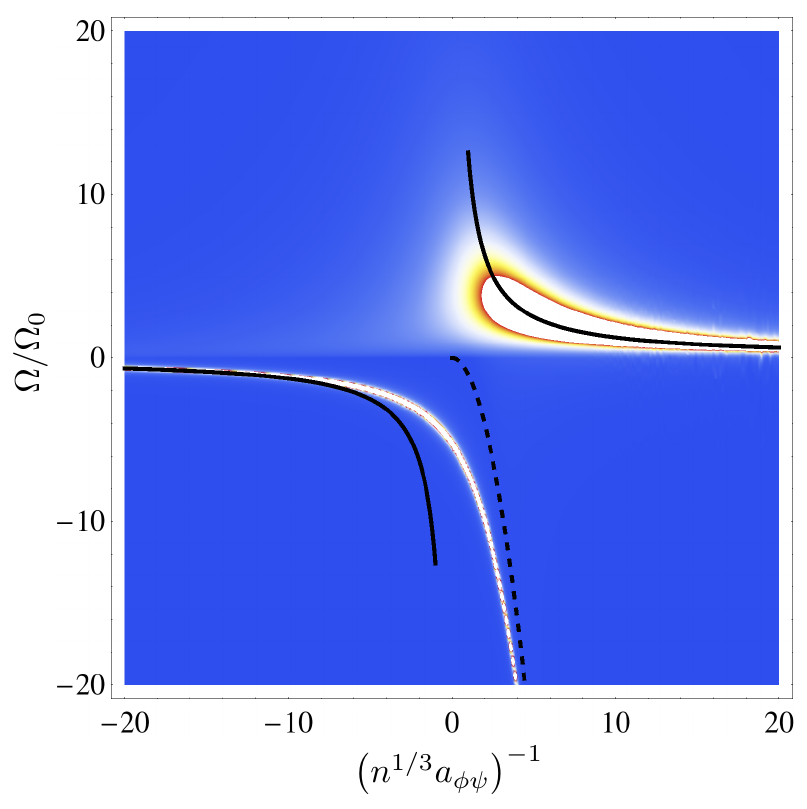}
  \caption{Impurity spectral function $A_\text{pol}(\Omega,\vec p = 0)$ in the
  non-selfconsistent T-matrix approximation with $n^{1/3}\aBB=0.1$. Solid lines:
  weak coupling limit (valid for both branches) $E\sim \gBF n$. Dashed line:
  energy of the universal dimer in vacuum, $\Edim\sim -1/\mB\aBF^2$. Here and in
  all following plots, zero-width peaks are given a small artificial width
  to be visible on the graph.}
  \label{fig:ApolgBB01}
\end{figure}

In Fig.~\ref{fig:ApolgBB01} we show the impurity spectral function calculated
using the \nsct approximation and evaluated at vanishing momentum as a function
of the dimensionless quantities $\Omega/\Omega_0$ and
$(n^{1/3}a_{\phi\psi})^{-1}$, where $\Omega_0=n^{2/3}/\mB$. The spectral
function shows a continuous background for positive frequencies $\Omega$
carrying little spectral weight and is dominated by two sharply peaked features,
one at positive and one at negative $\Omega$.  The sharp feature appearing at
low energies can be interpreted as an attractive polaron. It is a sharp
quasiparticle excitation giving rise to a delta peak in the spectral function
which carries most
of the spectral weight for large negative values of $\ainv$. In this weak
coupling regime we recover the perturbative result for the polaron's energy
which asymptotically obeys $\Eatt\sim\gBF n$.  Upon crossing the Feshbach
resonance where \ainv goes over to positive values, the attractive polaron
evolves smoothly into a molecular bound state which follows the energy of the
universal dimer, $\Eatt\sim\Edim=-\aBF^{-2}(\alpha+1)/2\alpha\mB$.

\begin{figure}[htbp]
  \centering
  \includegraphics{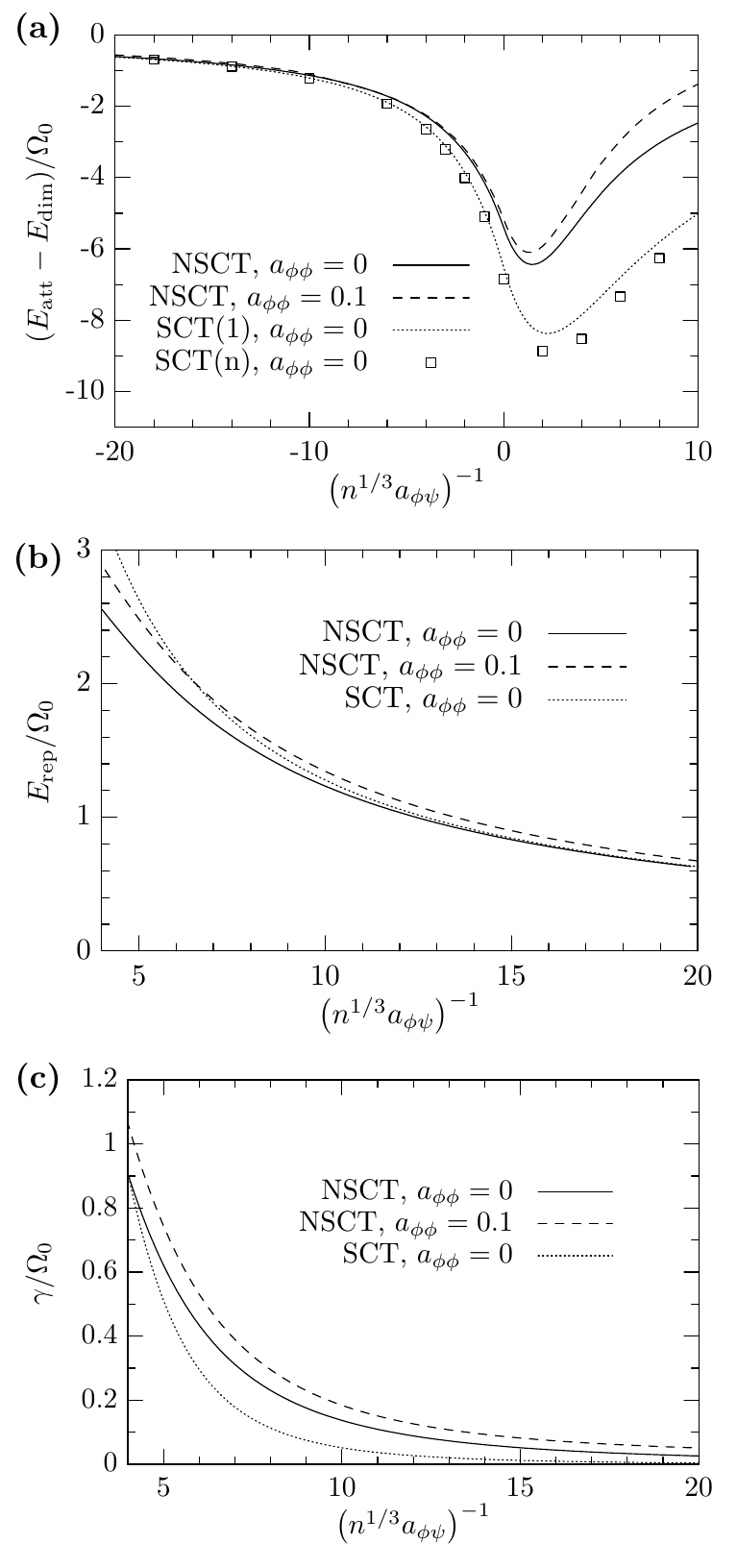}
  \caption{Quasiparticle energies and width obtained in the  non-selfconsistent
  T-matrix (\nsct) approach with both a finite and a vanishing boson--boson
  interaction, and the selfconsistent (\sct) approach.  In the latter case, the
  dotted line indicates the result after the first iteration of the
  selfconsistency loop while the individual dots denote the final value for
  $15$ iterations when full selfconsistency of the T-matrix equations is
  reached.  (a) Attractive polaron energy with the universal dimer binding
  energy
  $\Edim=-\theta(\aBF)/\mB\aBF^2$ subtracted.  This representation accentuates
  the absolute difference between the values obtained in the various
  approximation schemes. The relative deviation is far smaller on the right of
  the minimum due to the large modulus of \Edim.  (b) Energy of the repulsive
  polaron. The plot excludes low values of \ainv where the repulsive polaron is
  not properly defined. (c) Quasiparticle width of the repulsive polaron for the
  same range of values of \ainv.}
  \label{fig:energies}
\end{figure}

\begin{figure}[htbp]
  \centering
    \includegraphics{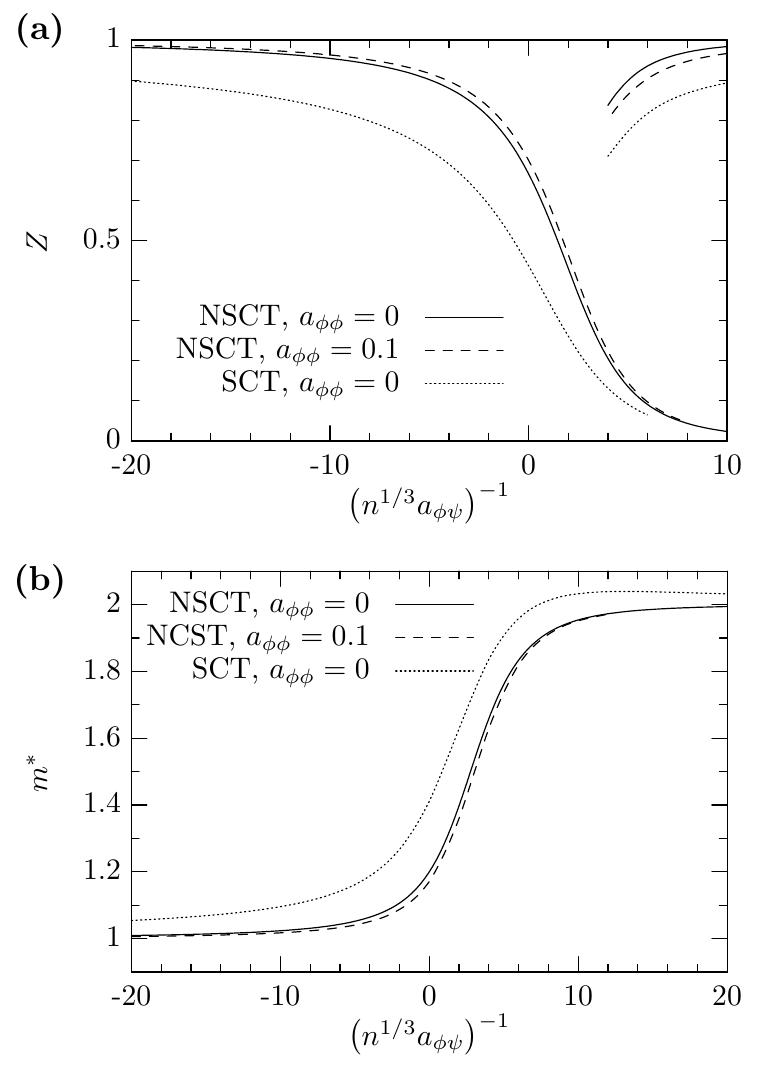}
  \caption{(a) Quasiparticle weights for both branches (the lines on the upper
  right of the graph correspond to the repulsive polaron). (b) Effective mass
  of the attractive polaron. Parameters and line styles are as in
  Fig.~\ref{fig:energies}. }
  \label{fig:ZandM}
\end{figure}

Although it remains the stable ground state, the attractive polaron loses
spectral weight in favor of a second feature which emerges at positive energies
[cf.~Figs.~\ref{fig:energies} and~\ref{fig:ZandM}]. This ``high-energy
excitation'' absorbs most of the spectral weight lost by the attractive polaron.
Its positive energy indicates that it corresponds to a quasiparticle interacting
with the Bose gas via an effective repulsion, hence it is coined the repulsive
Bose polaron in the following. The appearance of this repulsive branch  has
strong similarities to the ``upper branch'' present  in the case of a
two-component Fermi gas~\cite{jo2009,bart2011,shen2011} and in particular the
repulsive Fermi polaron \cite{cui2010,schm2011,mass2011}.  Similarly to the
latter case, the repulsive Bose polaron exists only at positive interspecies
coupling and becomes a well-defined quasiparticle only for moderate to small
values of the interaction constant \aBF as can be seen from its width $\gamma$
shown in Fig.~\ref{fig:energies}~(c).  The appearance of the repulsive polaron
suggests a simple physical picture where the impurity interacts
with the bosons via a positive scattering length, resulting in a positive
interaction energy. Indeed, we find that the energy of the repulsive polaron is
well approximated by the mean field result $\Erep\approx \gBF \rhoBEC$ wherever
the quasiparticle peak is well defined. This picture, however, neglects the
physical origin of the positivity of the scattering length which is the presence
of a bound state in the spectrum.  Indeed, as \aBF increases towards the
Feshbach resonance, the corresponding lifetime rapidly becomes so short that a
detailed experimental study of strongly repulsive polarons will be a major
challenge.

In Appendix~\ref{app:molecule_propagator} we argue that
when crossing the Feshbach resonance, there is actually a smooth crossover of
the ground state from a polaronic to a bound molecule state, a picture which
emerges naturally when considering the problem using a two-channel model instead
of the one-channel model discussed here. Essentially, this crossover, which
stands in clear contrast to the case of the Fermi
polaron~\cite{prok2008,prok2008b,punk2009,mora2009}, finds
its origin in a hybridization of the molecular and the polaronic state due to
the presence of the condensate~\cite{march2008}. This intuitive picture is
corroborated by the fact that the effective mass of the attractive polaron
crosses over from $\alpha\mB$, the mass of a single impurity, to $(\alpha+1)\mB$
which is the mass of a molecule made up of the impurity and one
boson~[cf.~Fig.~\ref{fig:ZandM}~(b)]. 

\begin{figure}[htbp]
  \centering
  \includegraphics{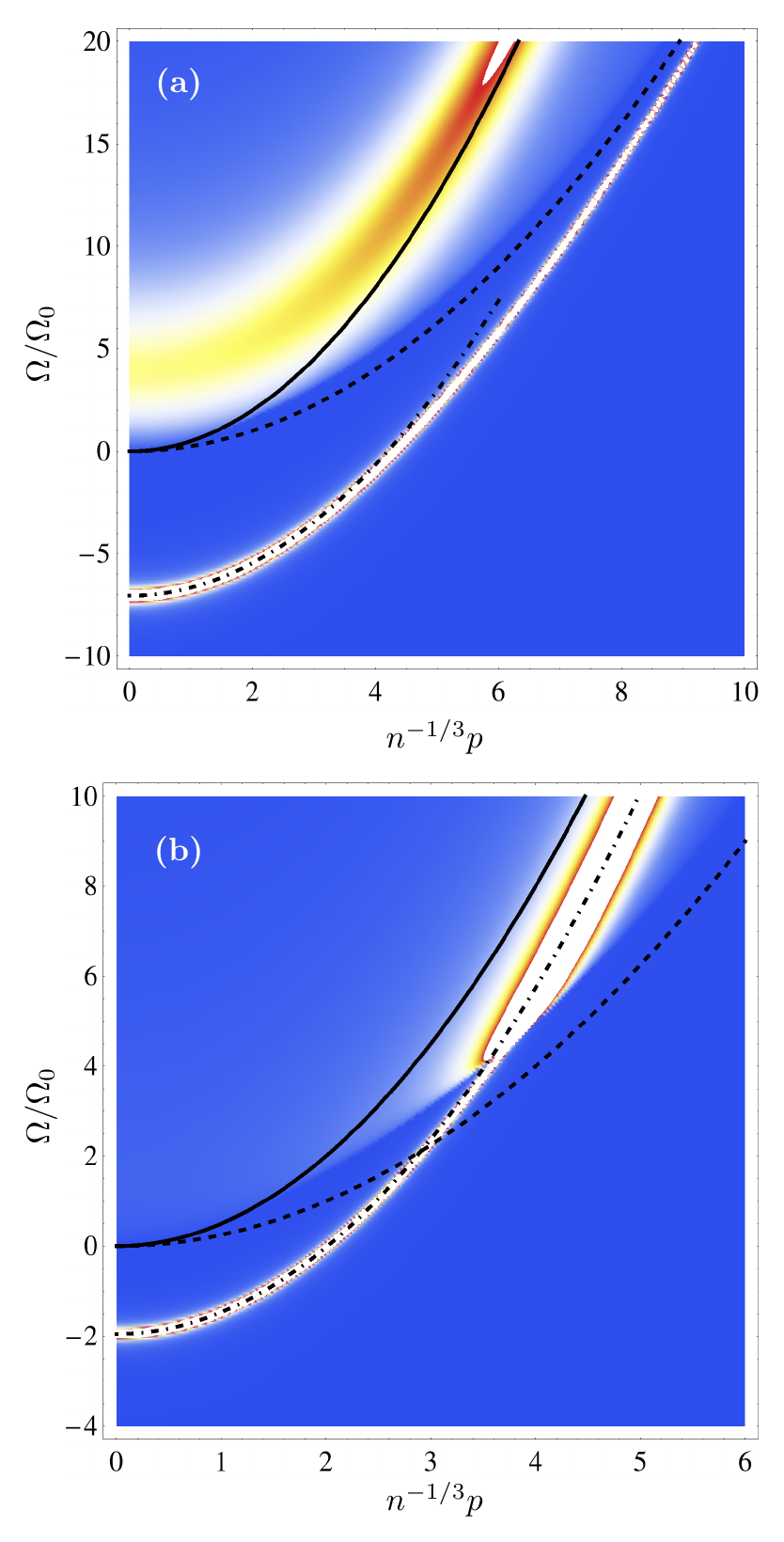}
  \caption{Impurity spectral function $A_\text{pol}(\Omega,\vec p)$ as a
  function of frequency and momentum for (a) $(n^{1/3}\aBF)^{-1}=1$ and (b)
  $(n^{1/3}\aBF)^{-1}=-5$.  In both graphs, $n^{1/3}\aBB=0.1$.
  Solid line: free impurity dispersion. Dashed line: free molecule like
  dispersion.  Dash-dotted line: dispersion according to the effective mass of
  the attractive polaron at $p=0$. For positive \aBF, the attractive polaron
  peak gradually bends away from its dispersion at vanishing momentum,
  reflecting an increase in the momentum-dependent effective mass. At negative
  \aBF, the effective mass stays approximately constant as a function of
  momentum.}
  \label{fig:Apolp}
\end{figure}

The behavior of the quasiparticles as a function of momentum, shown in
Fig.~\ref{fig:Apolp}, is qualitatively different on the two sides of the
Feshbach resonance. For $\aBF>0$, the attractive polaron's momentum dependent
effective mass increases with increasing momentum and its dispersion eventually
follows the dispersion of the molecular state which reflects the
polaron-to-molecule crossover also in the momentum domain. The onset of the
scattering continuum happens at $p^2/2\alpha\mB$ for $p/\alpha\mB\leq c=
\sqrt{\gBB n/\mB}$ (within the considered approximation, this statement is exact
and reflects Landau's critical velocity, cf.~Appendix~\ref{app:tmatrix}) and
crosses over to a molecule-like dispersion $\sim p^2/2(\alpha+1)\mB$ for larger
momenta.  The ``dispersion law'' of the continuum onset is determined by the
onset of the imaginary part of the pair propagator and is thus independent of
\aBF in the \nsct approach. As the attractive polaron peak approaches the
molecule dispersion, it loses more and more of its spectral weight in favor of
the repulsive polaron which gains spectral weight while its effective mass
approaches $\alpha\mB$. At large momenta the repulsive polaron becomes a free
particle as expected since at high momenta many-body fluctuations are suppressed
and the impurity becomes ignorant of its environment.  For negative \aBF the
repulsive polaron is absent~[cf.~Fig.~\ref{fig:Apolp}~(b)]. Here the attractive
polaron has a spectral weight close to one and an effective mass close to
$\alpha\mB$ for all momenta. As a consequence, it eventually crosses the
continuum onset and acquires a finite width. For large enough momenta, this
width becomes negligible and the attractive polaron behaves as a free particle.

Our results depend only weakly on the boson--boson coupling \gBB which suggests
to consider the simplifying limit $\gBB\to 0$. This limit should be understood
in the following sense: while a finite value of \gBB is essential for the
mechanical stability of the system, we are only interested in the effects of the
Bose gas on the impurity, not the inverse. The limit $\gBB\to 0$ is nothing but
a mathematical commodity which actually means that boson--boson interaction are
very weak but still sufficiently large to ensure
stability~\cite{yu2011,frat2012}. Carrying out the limit leads to important
simplifications in the calculation. First, all Bose propagators become free
particle propagators, thus eliminating the mathematically tricky Bogoliubov
dispersion. Second,  the loop diagram in Fig.~\ref{fig:impurity_diagrams}~(b)
vanishes identically, leaving only the mean field like contribution to the
impurity selfenergy~\eqref{eq:def_selfenergy}, i.e., the selfenergy is just
proportional to the T-matrix \tmatrix. Furthermore, all anomalous contributions
which are neglected in our calculation at finite $\gBB>0$ now indeed vanish. As
already stated, in the approximation $\gBB\to 0$ the in medium T-matrix is
identical to the T-matrix in vacuum given in Eq.~\eqref{eq:Lvacuum} up to the
impurity's chemical potential absorbed in $\Omega$. Considering the simplicity
of the selfenergy in the limit $\gBB\to 0$, it comes as a surprise that at
vanishing momentum the resulting spectral function turns out to be almost
identical to the one obtained for finite boson--boson
interactions~(cf.~Figs.~\ref{fig:energies} and \ref{fig:ZandM} for an explicit
comparison of the quasiparticle properties for finite and vanishing boson--boson
interactions). This is particularly remarkable since we deliberately chose the
rather large value $n^{1/3}\aBB=0.1$ for the plots involving a finite
boson--boson interaction.

The simplicity of the \nsct approximation with a
non-interacting BEC allows to give some simple analytical relations between quasiparticle properties. The first relates the quasiparticle weight and the effective mass.
In fact, since in this approximation the selfenergy is a function of
$\Omega-p^2/2(\alpha+1)\mB$, the effective mass can simply be expressed as
\begin{equation}
  \meff =
  \frac{1-\partial_\Omega\Re\self}
  {\frac{1}{\alpha}-\frac{1}{\alpha+1}\partial_\Omega\Re\self}
  \bigg|_{E(p)}
=
\frac{\alpha(\alpha+1)}{Z+\alpha}\ .
  \label{eq:mefffromZ}
\end{equation}
One sees that  the effective mass is strictly bounded by the impurity and the
molecule mass and follows a smooth
interpolation between the two as the scattering length is tuned across the
resonance. This nicely reflects the crossover of the polaron to a molecule
described in Appendix~\ref{app:molecule_propagator}.   The
quasiparticle weight $Z$ and the effective mass $\meff$ are shown as functions
of $\ainv$ in Fig.~\ref{fig:ZandM}. One may also determine the momentum at which
the attractive polaron enters the excitation continuum at small negative \aBF.
In this regime the polaron always has an effective mass close to $\alpha\mB$
such that  the polaron dispersion relation is well approximated by
$\Eatt(p)\approx\gBF n+p^2/2\alpha\mB$ while the onset of the continuum is
exactly at $p^2/2(\alpha+1)\mB$ (in contrast to the case of a non-zero \gBB
discussed above).  Consequently, the two intersect at
$p\approx\sqrt{-4\pi(\alpha+1)^2 n\aBF}$.  Finally, the energy of the attractive
polaron may be calculated analytically.  While the resulting expression is
cumbersome, it permits to obtain the leading correction to the universal dimer
energy which at small positive scattering lengths is given by $\Eatt\sim
(-1/\aBF^2-8\pi\aBF n)(\alpha+1)/2\alpha\mB$~\footnote{In terms of the
two-channel model discussed in Appendix~\ref{app:molecule_propagator}, the
second term may be interpreted as an effective boson--molecule repulsion.}.
\begin{figure}[htbp]
  \centering
    \includegraphics{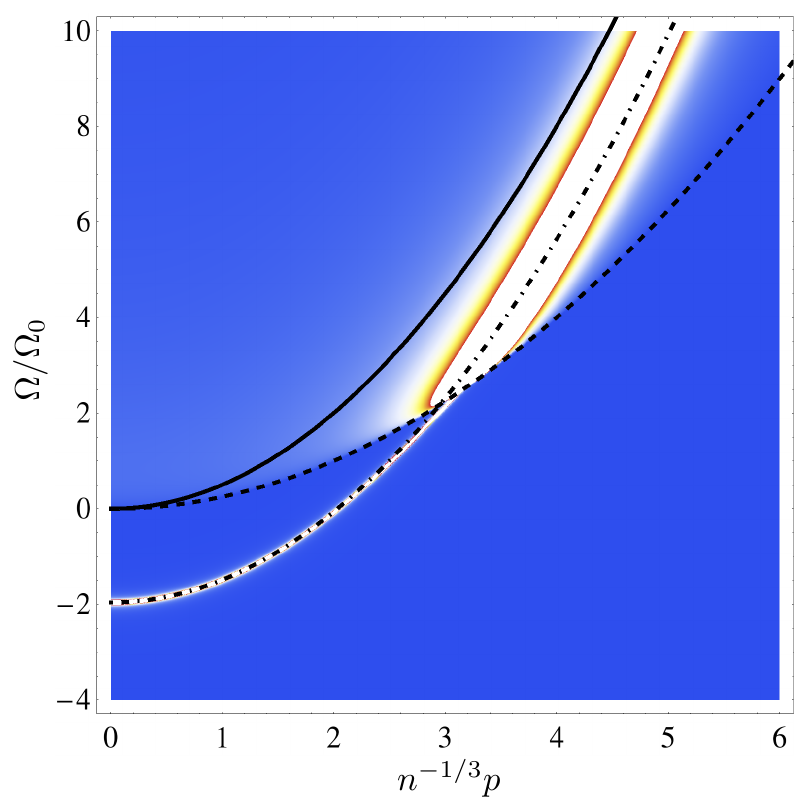}
    \caption{Spectral function $\Apol(\Omega,\vec{p})$ 
    for $\aBB=0$, all other parameters are as Fig.~\ref{fig:Apolp}~(b). Note
    that here the continuum onset coincides  with the dashed line indicating
    the free molecule like dispersion. The point where the polaron peak
    acquires a finite width is shifted accordingly.}
  \label{fig:ApolpgBB0}
\end{figure}

The difference in the ``dispersion'' of the continuum onset depending on whether
one assumes a finite \gBB or not is in fact the only qualitative difference that
can be seen in the spectral function~[cf.~Fig.~\ref{fig:ApolpgBB0}, where, in
contrast to Fig.~\ref{fig:Apolp}~(b), the continuum onset coincides with the
free molecule dispersion marked by the dashed line]. As a consequence, while
there is a visible deviation in the crossing between the attractive polaron and
the continuum onset at positive \aBF, the relation~\eqref{eq:mefffromZ} between
$Z$ and \meff is satisfied to a good approximation even for $\gBB>0$.  To
summarize, within the chosen approximation scheme the influence of \gBB,  in
particular on quantities evaluated at vanishing momentum, is almost invisible in
spite of the very large value $n^{1/3}\aBB=0.1$ used in the data shown before.

\section{Selfconsistent T-matrix approximation}
\label{sec:SCT}

In the \nsct approach the backaction of the impurity selfenergy on the in-medium
T-matrix is neglected since only bare impurity Green's functions appear on the
right-hand side of the
T-matrix equation depicted in Fig.~\ref{fig:impurity_diagrams}~(c). In this
Section we include this feedback by solving the Bose polaron problem using a
\textit{selfconsistent} T-matrix (\sct) approach where the thin impurity line on
the right-hand side of Fig.~\ref{fig:impurity_diagrams}~(c) becomes bold, i.e.,
the bare impurity propagator is replaced by its dressed counterpart determined
by Dyson's equation. Motivated by our observation that within our T-matrix
approach the inclusion of a finite boson--boson repulsion yields only
quantitative corrections to the results~\footnote{We have also compared some of
the data discussed here to results obtained in a hybrid selfconsistent
approximation where one uses the T-matrix and boson propagator for an
interacting condensate, but neglects the one-loop correction to the selfenergy.
We find that the differences between a finite and a vanishing \gBB are even
smaller in the selfconsistent approximation.}, we only consider the limit
$\gBB\to 0$ in the following.

In the context of ultracold gases in the continuum, the selfconsistent T-matrix
approach has been employed  by Haussmann and coworkers for the study of the
BEC--BCS crossover both for thermodynamic and dynamical quantities
\cite{haus2007,haus2009,enss2011,enss2012a}.  For instance, their prediction for
the equation of state is in very good agreement with experiments~\cite{ku2012}
and state of the art bold diagrammatic Monte-Carlo calculations~\cite{houc2012}
in both the superfluid and the normal phase of the balanced Fermi gas.   As we
will see, the inclusion of selfconsistency leads to important quantitative
changes in the ground state properties as well as qualitative changes in the
excitation spectrum.

The inclusion of the dressed impurity Green's function
leads to the modified many-body T-matrix equation 
\begin{multline}
\frac{1}{\tmatrix(i\omega,\vec{p})}=\ginv + \int_{\vec{k}}\Bigg\{ 
\int_{\nu}G_\phi^{(0)}[i( \omega-\nu),\vec{p}-\vec{k}]
  \\
  \times\frac{1}{i\nu-\frac{\vec{k}^2}{2\alpha}+\muF-\rhoBEC\tmatrix(i\nu,\vec{k})} 
  -\frac{2\alpha \mB}{(\alpha+1)\vec{k}^2}
  \Bigg\}\ .
  \label{eq:sc_tmatrix}
\end{multline}
In contrast to the \nsct approach, this equation includes the virtual excitation
of an arbitrary number of bosons from the condensate.  Unlike in
Eq.~\eqref{eq:tmatrix_eqn}, the unknown function $\tmatrix(i\omega,\vec p)$ now
appears (in a non-linear way) on both sides of the integral equation which
complicates the calculation significantly.  We perform this calculation using
imaginary frequencies because the resulting Green's functions are smooth
functions of the frequency, which makes them easier to handle than their
counterparts at real frequencies---the latter  may develop nonanalytic features
such as kinks as we have seen in the previous Section.  While the angular part
of the momentum integration can be performed analytically, there is no evident
way to do so for the frequency integration because the analytical structure of
the T-matrix \tmatrix on the right-hand side is a priori unknown. This leaves us
with two integrals that have to be evaluated numerically.  

The mathematical structure of Eq.~\eqref{eq:sc_tmatrix} resembles the
renormalization group equations obtained for the pair propagator in the study of
the Fermi polaron using the functional renormalization group \cite{schm2011}.
Thus, we may use a similar numerical scheme to calculate the right-hand side of
this equation.  To this end, we discretize the T-matrix on a grid in momentum
and frequency space $(i\omega_k,p_l)$.  Due to spatial isotropy, the grid
depends only on frequencies and absolute momenta. Furthermore, we exploit the
fact that for high frequencies and momenta the many-body T-matrix reduces to the
T-matrix in vacuum to a very good approximation.  The maximum extent of the
resulting finite grid  is chosen such that the error due to this truncation is
smaller than the accuracy of the final results. The latter is limited by the
finite number of grid points as well as the finite number of steps used to solve
Eq.~\eqref{eq:sc_tmatrix} iteratively. The chosen grid contains
$\mathcal{O}(10^3)$ grid points.  The frequency and momentum integration in
Eq.~\eqref{eq:sc_tmatrix} is performed using a bicubic spline interpolation of
the numerical data as done in \cite{schm2011}. 

In order to solve Eq.~\eqref{eq:sc_tmatrix} selfconsistently we rely on an
iterative scheme. We start by solving this equation for an initial choice
$\Gamma^{(0)}_{\phi\psi}$ of the T-matrix appearing on the right-hand side given
by the non-selfconsistent T-matrix obtained in the previous Section.  The
evaluation of the right-hand side then yields an ``improved'' T-matrix
$\Gamma_{\phi\psi}^{(1)}$ which in turn is reinserted into the right-hand side
of Eq.~\eqref{eq:sc_tmatrix}. This yields the next iteration
$\Gamma^{(2)}_{\phi\psi}$ and so on. The process is iterated until convergence
to a numerically stable result is reached. We find that $n\lesssim 10$
iterations are typically sufficient. To verify the numerical stability of the
final result, however, we carry on until up to $n=20$ iterations.

From the T-matrix $\Gamma_{\phi\psi}^{(n)}$ in the $n$th step of this
``selfconsistency loop'' we calculate the improved, dressed impurity Green's
function  $G_\psi^{(n)}$ via Dyson's equation depicted in
Fig.~\ref{fig:impurity_diagrams}~(a). In the numerical solution using imaginary
frequencies, it is of utmost importance to adjust the impurity chemical
potential $\mu_\psi$ to a value that guarantees that the impurity atom does not
have a finite occupation in the final iteration, i.e., one has to satisfy the
condition
\begin{multline}
  \left[G_\psi^{(0)}(i\omega=0,\vec p;\mu_\psi)\right]^{-1}-\rho_0
  \Gamma_{\phi\psi}^{(n)}(i\omega=0,\vec p;\mu_\psi)\leq 0
\label{eq:condition}
\end{multline}
for all $\vec p$. In fact, if we require Eq.~\eqref{eq:condition} to be an
equality with the choice of $\muF=\mu_\text{pol}$ at $\vec{p}=0$, this uniquely
determines the ground state energy of the attractive polaron via the basic
definition of the chemical potential, $\mu_\text{pol}=E(N_\psi+1)-E(N_\psi)$,
where $N_\psi=1$ for the impurity problem, hence $E_\text{att}=\mu_\text{pol}$.
The energies obtained from the selfconsistent T-matrix approach are shown in
Fig.~\ref{fig:energies} along with the results of the \nsct approximation.
Specifically at unitarity where changes with respect to the \nsct schema are
particularly strong, we find $E_\text{att}/\Omega_0=-6.84(1)$ while the \nsct
approach yields $E_\text{att}/\Omega_0=-5.390$.  We see that the changes are of
the order of 25 percent around unitarity.

To study the impact of selfconsistency on the excitation spectrum of the system,
we have to perform an analytical continuation of the impurity Green's function
to real frequencies. Although this is in principle possible, it would require
the analytical continuation from a finite, discrete set of numerical data which
is a mathematically ill-defined problem. In order to avoid the related
complications and yet obtain fairly precise results, we make use of the
observation that the by far largest effect of selfconsistency arises already in
the first iteration [cf.~the dotted line in Fig.~\ref{fig:energies}~(a)].  Thus
we expect that all major aspects of the impact of selfconsistency on the
impurity spectral function are already present at this stage. The numerical cost
of this first iteration is modest. One only has to calculate the integral
appearing in Eq.~\eqref{eq:pair_propgator}, but with the bare Fermi propagator
replaced by
\begin{equation}
  \GFhalf(\Omega,\vec{p})=
  \frac{1}{\Omega-\frac{\vec{p}^2}{2\mF}-\rhoBEC\Gamma_{\phi\psi}^{\text{NSC}}
  (\Omega,\vec{p})+i0^+}
  \label{eq:GFhalf}
\end{equation}
where $\Gamma_{\phi\psi}^{\text{NSC}}$ is the non-selfconsistent T-matrix we
obtained in the previous Section, cf.~Eqs.~\eqref{eq:tmatrix_eqn} and
\eqref{eq:pair_propgator}. Since the integrand is known analytically, the
frequency integration and subsequent continuation to real frequencies
may be carried out analytically so that only the momentum integral has to be
done numerically.

\begin{figure}[htbp]
  \centering
    \includegraphics[width=\linewidth]{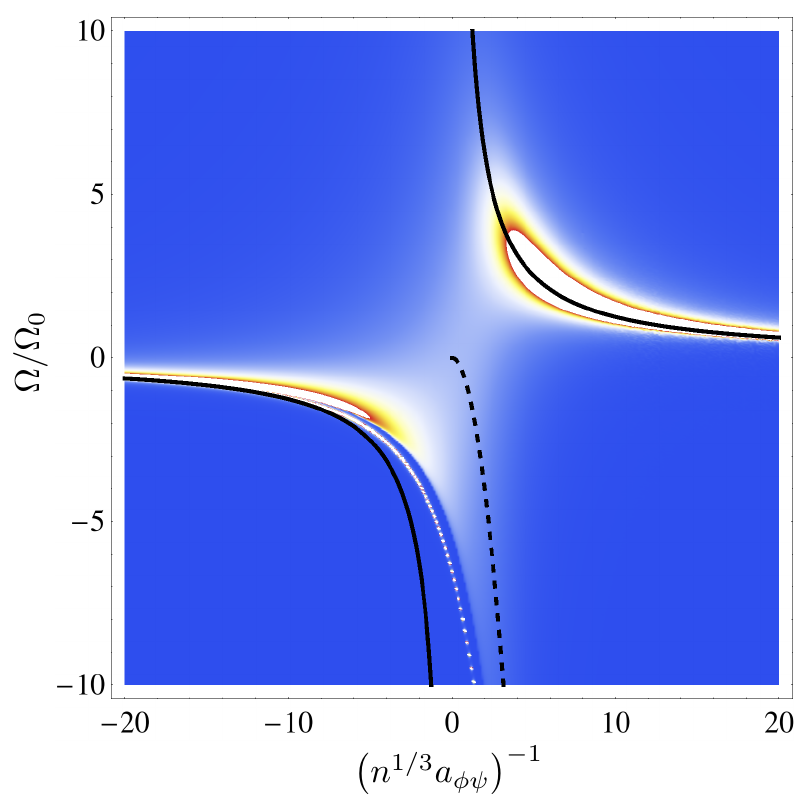}
  \caption{Polaron spectral function calculated using the 
  selfconsistent scheme discussed in Section~\ref{sec:SCT}. The black lines are
  identical to the ones in Fig.~\ref{fig:ApolgBB01}. Note how the continuum of
  excitations now follows the attractive polaron peak.}
  \label{fig:Apolsc}
\end{figure}

The qualitative changes following from this scheme with respect to the
non-selfconsistent one can be seen in Fig.~\ref{fig:Apolsc}.  The continuum
onset is pulled to negative energies and now follows the curve which in the
non-selfconsistent approximation is described by the polaron peak. A similar
behavior has been found for the molecule spectral function in a renormalization
group study of the Fermi polaron~\cite{schm2011}. This may be seen as a further
indication of the Bose polaron being hybridized with the molecule,
cf.~Appendix~\ref{app:molecule_propagator}.  Moreover, we observe a strong
suppression of both the attractive and the repulsive polaron's spectral weight.
This large suppression, however, does not come as a surprise. Indeed, by solving
Eq.~\eqref{eq:sc_tmatrix} selfconsistently we incorporate more fluctuations
which entangle the impurity with the bosons' degrees of freedom. This loss of
spectral weight in the attractive polaron is compensated by a transfer of weight
to excited continuum states which is facilitated by the reduced excitation gap
as compared to the \nsct approach. The substantial reduction of the
quasiparticle weight is accompanied by changes to the effective mass of a
similar magnitude. In the selfconsistent calculation the effective mass is no
longer related to the quasiparticle weight by the simple
relation~\eqref{eq:mefffromZ} and takes values larger than $(\alpha+1)\mB$.
However, apart from a small upward shift, it still essentially follows its
behavior from the \nsct approximation [cf.~Fig.~\ref{fig:ZandM}~(b)].  The
repulsive polaron is shifted to slightly higher
energies~[cf.~Fig.~\ref{fig:energies}] while its quasiparticle width is
substantially reduced with respect to the \nsct result for all but the strongest
interspecies interactions.

\begin{figure}[htbp]
  \centering 
    \includegraphics{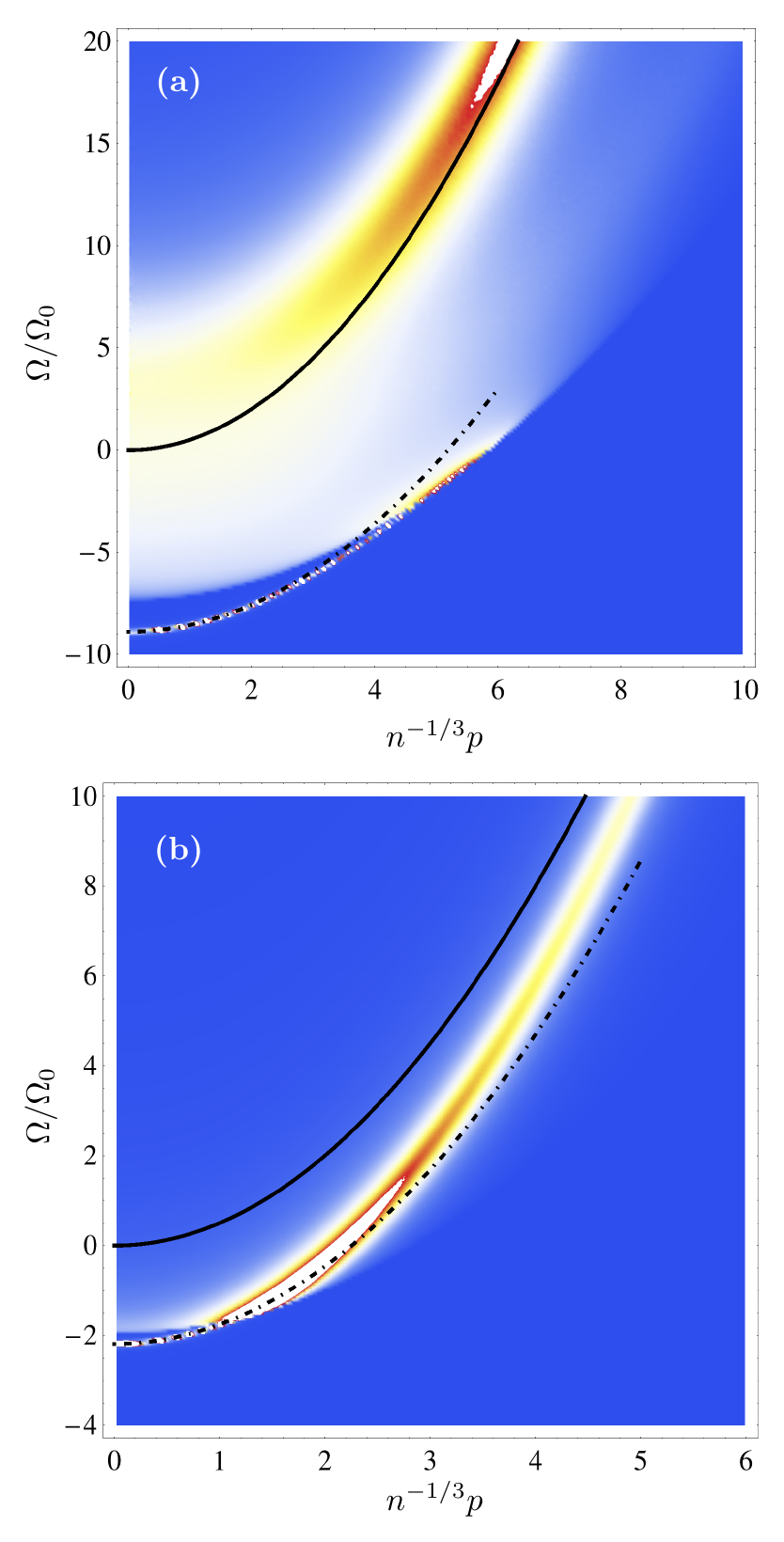}
  \caption{Polaron spectral function calculated using the 
  selfconsistent scheme discussed in Section~\ref{sec:SCT} as a function of
  momentum and frequency. The black lines have the same meaning as in
  Fig.~\ref{fig:Apolp}. (a) $(n^{1/3}\aBF)^{-1}=1$. Note how the attractive
  polaron peak is ``annihilated'' when it touches the continuum. (b) $(n^{1/3}\aBF)^{-1}=-5$.
  The continuum practically merges with the peak for higher momenta.}
  \label{fig:Apolscp}
\end{figure}

Further changes can be seen when one considers the momentum dependence of the
impurity spectral function which is shown in Fig.~\ref{fig:Apolscp} for two
different values of the interspecies coupling. The most prominent differences
appears for positive values of \ainv [Fig.~\ref{fig:Apolscp}~(a)] where the
attractive polaron peak touches the continuum and dies out rapidly instead of
running parallel to it as it does in the \nsct approximation. The qualitative
behavior of the repulsive polaron, however, remains unchanged, showing a smooth
interpolation from a very broad peak at low momenta to a sharp peak with an
effective mass of $\alpha\mB$ towards higher momenta.  The changes with respect
to the \nsct approach are less pronounced for negative values of \ainv
[Fig.~\ref{fig:Apolscp}~(b)]. Here one only notices that when the attractive
polaron pole enters the continuum---which happens at lower momenta than in the
\nsct approximation due to the continuum's being pulled to negative
frequencies---the latter is gradually absorbed by the former. Hence, within the
\sct approach, the attractive polaron becomes subject to damping above a
critical momentum for \emph{any} interspecies interaction strength.

\section{Conclusion and Outlook}
\label{sec:discussion}

We have determined the excitation spectrum of an impurity immersed in a
homogeneous BEC.  We find that this spectrum is dominated by two branches, the
attractive and repulsive Bose polaron. The attractive polaron is a stable
quasiparticle at negative energies which exists for all interspecies couplings.
It exhibits a crossover from a weakly dressed polaron to a molecule as one
crosses the Feshbach resonance. This can be understood in terms of its
hybridization with a molecular state due to the BEC. The repulsive polaron
emerges as a metastable quasiparticle on the $\ainv>0$ side of the Feshbach
resonance. While it is long-lived for small scattering lengths \aBF, its
lifetime becomes exceedingly small as the Feshbach resonance is approached.  
The essence of the problem can already be captured by
a non-selfconsistent calculation using a vanishing boson--boson coupling
constant. The most important correction to this simple picture is given by the
multiple excitation of bosons from the condensate, which we account for by the
selfconsistent incorporation of the selfenergy feedback into the T-matrix
equation. We predict various quasiparticle properties of the attractive and
repulsive polaron which can be tested in future experiments. For instance,
radiofrequency experiments permit to measure not only the excitation
energies~\cite{schi2009}, but also the quasiparticle weight and width. The
latter two can be inferred from a shift in the Rabi frequency and the damping of
Rabi oscillations, respectively~\cite{kohs2012}.  The effective mass may in turn
be determined using momentum-resolved photoemission or Raman
spectroscopy~\cite{dao2007,stew2008,feld2011,korr2011}.

It is an interesting question what happens on time scales longer than those
relevant for inverse rf experiments.  For instance, the repulsive polaron
exhibits a positive energy which decreases when $n\aBF^3$ is lowered. Thus, in
an experimental situation with a trapped BEC, the repulsive polaron can minimize
its energy by moving to a region of lower density~\cite{wu2011}. The fact that
this tendency towards phase separation happens for any positive \aBF is in stark
contrast to the case of the repulsive Fermi polaron. In the latter, phase
separation---which here may also be seen as a transition towards a ferromagnetic
phase---only happens for interaction strengths above a critical value due to the
competition between kinetic and interaction energy~\cite{schm2011,mass2011}.
For the Bose polaron, the process of phase separation is itself in
competition with the tendency towards self-localization accompanied by a local
deformation of the BEC. Concerning the study of such dynamical phenomena, our
calculation can be seen as the derivation of an effective field theory for the
repulsive Bose polaron which includes quasiparticle properties such as a finite
lifetime and an effective momentum-dependent interaction.  Starting from the
corresponding equations of motion, the time evolution from the initial
out-of-equilibrium state towards self-localization or phase separation may now
be studied.  Note that our discussion so far ignores the effects of three-body
recombination due to Efimov physics~\cite{efim1970}. Even in our inverse rf
spectroscopy scenario, the latter is not completely suppressed.  In the case of
the repulsive polaron, one may however even utilize the Efimov effect to
suppress losses by exploiting the minima in the three-body recombination which
are due to the destructive interference of decay channels~\cite{helf2010}. In
fact, it poses an interesting question on its own how Efimov physics is affected
by medium effects~\cite{zinn2013} such as the hybridization of the impurity with
the molecular state.

Finally, it would be interesting to investigate the possibility of an
alternative representation of the Bose polaron.  In the Fermi polaron problem
the NSCT approach leads to equations which are formally equivalent to the
equations obtained from a simple variational wave function
ansatz~\cite{chev2006a}.  This ansatz describes the  Fermi polaron as a bare
impurity dressed by a single particle--hole fluctuation.  We expect that such a
mapping from diagrams to a variational wave function exists as well in the
present case of the Bose polaron. 

\begin{acknowledgements}
We thank Wilhelm Zwerger for suggesting the problem and for careful reading of
this manuscript. We also acknowledge helpful discussions with Marcus Barth,
Eugene Demler, Tilman Enss, Francesco Piazza and Alessio Recati. This work has
been supported by the DFG Forschergruppe~801.
\end{acknowledgements}

\appendix

\section{Hybridization, molecule Green's function and polaron-to-molecule crossover}
\label{app:molecule_propagator}

In experiments, large scattering lengths can be achieved by the use of Feshbach
resonances. Here the dependence of the energy level of a closed-channel molecule
on an external magnetic field is exploited to obtain arbitrarily large
scattering lengths when the state approaches the open-channel scattering
threshold. The basic physics of a Feshbach resonance can be described by the use
of a simple two-channel model which explicitly includes the molecular state as a
dynamical degree of freedom. While detailed accounts on the physics of the
two-channel model in the context of Fermi- and Bose gases can be found in
literature \cite{gora2004,szym2005,bloc2008,wern2009,chin2010}, we will focus here on the particularities which arise in
the mixed Bose--impurity system due to the presence of the condensate. In
particular, the study of the two-channel model allows to connect our results for
the T-matrix $\Gamma_{\phi\psi}$, obtained in the single-channel model employed so far, with the propagator of the molecule. In this
context we highlight the differences in
comparison to the case of a Fermi polaron.  Specifically, in a description of
the Fermi polaron in the limit of open-channel dominated Feshbach resonances,
the T-matrix can be interpreted as the molecule propagator when physics close to
the resonance is considered.  The spectral function of the molecule is then
simply proportional to the imaginary part of the T-matrix. While in the Fermi
polaron problem this identification even allows for a qualitatively correct
description of the polaron-to-molecule transition present in this system
\cite{comb2007,comb2009,punk2009}, we show in this Appendix how such a simple identification of
the T-matrix with the molecule propagator fails in the context of the Bose
polaron where no transition but rather a crossover occurs.

The two-channel model describing a  Bose
polaron close to a Feshbach resonance of arbitrary width \cite{chin2010}  is given
by
\begin{multline}
  S=\int\d\tau\int\d^3x\,\left\{ 
  \varphi^*_{\vec{x},\tau}\left( \partial_\tau-\frac{1}{2\mB}\nabla^2-\mu_\phi \right)
  \varphi_{\vec{x},\tau}
  \right. \\ 
+\psi^*_{\vec{x},\tau}\left( \partial_\tau-\frac{1}{2\alpha \mB}\nabla^2-\mu_\psi \right)
  \psi_{\vec{x},\tau}
  \\ \left.
+\xi^*_{\vec{x},\tau}\left( \partial_\tau-\frac{1}{2(\alpha+1)\mB}\nabla^2+\nuM \right)
  \xi_{\vec{x},\tau}
  \right. \\  \left.
  +\frac{\gBB}{2}|\varphi_{\vec{x},\tau}|^4
  +h\left(\xi^*_{\vec{x},\tau}\psi_{\vec{x},\tau}\varphi_{\vec{x},\tau}
  +\psi^*_{\vec{x},\tau}\varphi^*_{\vec{x},\tau}\xi_{\vec{x},\tau}\right)
  \right\}\  ,
  \label{eq:two_channel_action}
\end{multline}
where the field $\xi$ describes the molecular state in the closed channel 
which is a composite state of an impurity and a boson.  The interaction between
the impurity $\psi$ and the bosons $\varphi$ is mediated by the exchange of a
molecule $\xi$ which couples via  the conversion coupling $h$ to the
impurity and the bosons. The analysis of the few-body physics governed by
Eq.~\eqref{eq:two_channel_action} shows that $h$ is connected to the
experimentally accessible characteristic width of the Feshbach resonance $r^*$
via $h^2\sim1/r^*$. While the limit $h\to\infty$ describes so-called
open-channel dominated Feshbach resonances, a resonance is termed
closed-channel dominated in the limit $h\to 0$.

The molecule is a physically propagating degree of freedom and as such is
supplemented by a dynamical propagator already on the level of the classical
action \eqref{eq:two_channel_action}. The parameter $\nuM$ denotes the detuning
of the energy of the bare molecular state with respect to the atom scattering
threshold and depends on the external magnetic field \cite{gora2004,szym2005,bloc2008,wern2009,chin2010}.
The action \eqref{eq:two_channel_action} is quadratic in the field $\xi$, and by
integrating out $\xi$ in the path integral one easily verifies that the
two-channel model is equivalent to the single-channel model
\eqref{eq:microscopic_action} in the open-channel dominated limit
where $h\to\infty$. In this limit the momentum dependence of $\xi$ becomes
irrelevant and one finds the identification $\gBFtilde=-h^2/\nuM$. 
 
If we treat the bosons in the Bogoliubov approximation and neglect the
backaction of the impurity on the condensate, we obtain
\begin{widetext}
\begin{multline}
  S_{\text{eff}}=\int_{p}\left\{ 
  \frac{1}{2}
  \begin{pmatrix}
    \phi^*_p \\ \phi_{-p}
  \end{pmatrix}
  \begin{pmatrix}
    -\left[\GBbare(-p)\right]^{-1} & \gBB\rhoBEC \\
    \gBB\rhoBEC & -\left[\GBbare(p)\right]^{-1}
  \end{pmatrix}
  \begin{pmatrix}
    \phi_p \\ \phi^*_{-p}
  \end{pmatrix}
  +
  \begin{pmatrix}
    \psi^*_p \\ \xi^*_p
  \end{pmatrix}
  \begin{pmatrix}
    -\left[\GFbare(p)\right]^{-1} & h\sqrt{\rhoBEC} \\
    h\sqrt{\rhoBEC} & -\left[\GMbare(p)\right]^{-1}
  \end{pmatrix}
  \begin{pmatrix}
    \psi_p \\ \xi_p
  \end{pmatrix}
  \right\}
  \\
  +h\int_x\left(\xi^*_x\psi_x\phi_x+\psi^*_x\phi^*_x\xi_x  \right)\ ,
  \label{eq:two_channel_Bogoliubov}
\end{multline}
\end{widetext}
where \GFbare is the
bare impurity propagator as used in the main body of the article, and $\GMbare =
[i\omega-\vec{p}^2/2(\alpha+1)\mB-\nuM]^{-1}$ is the bare molecule propagator.
From this form of the action, one may directly read off the mean field
(fermionic) propagators
\begin{equation}
  \begin{split}
    \GFfull(\Omega,\vec{p}) =\frac{\GFbare(\Omega,\vec{p})}
    {1-h^2\rhoBEC\GMbare(\Omega,\vec{p})\GFbare(\Omega,\vec{p})}
    \\
    \GM(\Omega,\vec{p}) = \frac{\GMbare(\Omega,\vec{p})}
    {1-h^2\rhoBEC\GMbare(\Omega,\vec{p})\GFbare(\Omega,\vec{p})}
  \end{split}\ .
  \label{eq:hybridized_propagators}
\end{equation}
In the limit $\rhoBEC\rightarrow 0$, i.e., in absence of a condensate, these
propagators reduce to  their usual bare form also encountered for instance in
the context of the Fermi polaron close to a narrow Feshbach resonance
\cite{kohs2012, mass2012,schm2013}. In the presence of a BEC, however, the
propagators are hybridized~\cite{march2008} which constitutes a fundamental
difference between pure Fermi--Fermi and Bose--Fermi mixtures already on the
classical (mean field) level.

\begin{figure}[htbp]
  \centering
    \includegraphics{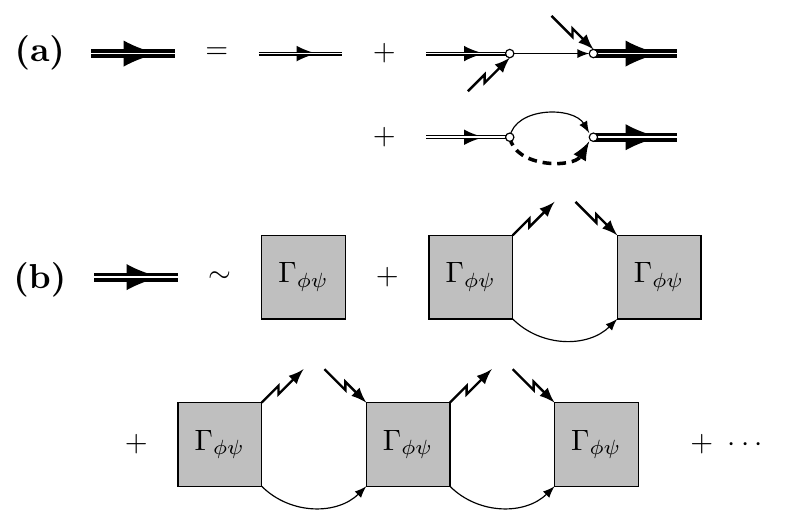} 
  \caption{(a) Dyson's equation for the molecule propagator in the two-channel
  model \eqref{eq:two_channel_Bogoliubov} in a T-matrix like approximation.
  Double lines denote the molecule propagator and open circles represent the
  atom--molecule conversion coupling $h$. (b) Diagrammatic representation of the
  molecule propagator within a single-channel description of the Bose polaron.
  Note that in order to establish the correspondence
  $\mathcal{G}_\xi\leftrightarrow\tmatrix$, one has to assume implicit
  conversion coupling vertices at both ends of the propagator line on the 
  left-hand side.}
  \label{fig:mol_fullprop}
\end{figure}

While the results of the NSCT approximation presented in Section~\ref{sec:nsct}
can be recovered in the limit $h\to\infty$, the
two-channel model also allows to generalize these results to resonances of
arbitrary width. Although this poses an interesting problem on its own, we
focus here on the question of how the presence of hybridization alters the
identification of the molecule with the T-matrix which is simply a
proportionality in the Fermi polaron problem close to a broad Feshbach resonance
\cite{comb2009,mora2009,punk2009,zoel2011,mass2012,schm2012,schm2013}.
To study this question, we calculate the molecular Green's function
$\mathcal{G}_\xi$ which is renormalized by the ladder diagram  depicted in
Fig.~\ref{fig:mol_fullprop}~(a). This leads to
\begin{multline}\label{eq:fullmolprop}
[\mathcal{G}_\xi(\Omega,\vec p)]^{-1}=\Omega-\frac{\vec{p}^2}{2(\alpha+1)\mB}\\
- h^2\rho_0 \GFbare(\Omega,\vec{p}) + h^2\Gamma_{\phi\psi}^{-1}(\Omega,\vec{p})\ ,
\end{multline}
where $\Gamma_{\phi\psi}$ appears upon the evaluation of the particle-particle loop diagram depicted in Fig.~\ref{fig:mol_fullprop}~(a).
We now take the limit
of an open-channel Feshbach resonance $h\to\infty$. This results in the
irrelevance of the dynamical part  in the first line of
Eq.~\eqref{eq:fullmolprop}  and $\mathcal{G}_\xi$ reduces
to
\begin{equation}
  h^2\mathcal{G}_\xi(\Omega,\vec{p}) = \frac{\tmatrix(\Omega,\vec{p})}
  {1-\rhoBEC\tmatrix(\Omega,\vec{p})\GFbare(\Omega,\vec{p})}\ .
  \label{eq:Gxi}
\end{equation}
From this expression the correspondence between $\tmatrix$ and the molecule
propagator within a single-channel description becomes apparent. It is given by
the resummation of the diagrammatic series depicted in
Fig.~\ref{fig:mol_fullprop}~(b). This  diagrammatic
structure does not exist in the Fermi polaron problem as it is solely a
consequence of the hybridization originating from the presence of the
condensate. Note that the expression on the right-hand side of
Eq.~\eqref{eq:Gxi} is precisely the quantity called $T$ (as opposed to $\Gamma$)
by the authors of~\cite{wata2008,sogo2013}. Hence, their distinction between
these two quantities may also be seen as one between the T-matrix and the
molecule propagator.

The derivation of the spectral function of the propagating degrees of freedom
necessitates the diagonalization of the full \textit{matrix} Green's function
which mixes the fermionic and molecular states in the Nambu spinor
$\chi=(\psi,\xi)^T$. In consequence the resulting full Green's functions for the
impurity and the molecule exhibit the same pole structure which is the essence
of the hybridization of the molecule with the impurity degree of freedom.  The
inspection of the denominator in~\eqref{eq:Gxi} now reveals that it is identical
to the denominator of the impurity Green's function $\GFhalf$ in
Eq.~\eqref{eq:GFhalf} which is obtained in the approximation which only takes 
into account the repeated impurity--condensate scattering. Hence, the molecule
and the polaron have indeed the same pole structure and we have successfully
established a consistent relation between the molecule propagator and the
T-matrix in a single-channel description. 

The description in terms of the Nambu spinor $\chi$ also allows to characterize
the nature of the eigenstates with respect to which the corresponding matrix
propagator becomes diagonal in field space. This diagonalization is achieved
using a transformation matrix $U(\theta)\in{\text{SO}}(2)$, where
$\theta\in[0,\pi/2]$ is a scattering length dependent mixing angle. For
$\theta=0$, the matrix Green's function is diagonal and the fields $\psi$ and
$\xi$ are its eigenstates. For finite values of $\theta$, however, the
eigenstates are superpositions of the two fields. In dependence on the
scattering length, the tangent $ \tan(2\theta)$ falls off to zero for weak
attraction while it has a pole close to unitarity. As a consequence of this
pole, the eigenstate which starts as the impurity state on the side of small
negative \aBF ends up as the molecular state on the side of small positive \aBF
and vice versa, in agreement with the intuitive picture of a smooth
polaron-to-molecule crossover.

\section{Regularization of the boson--boson interaction}
\label{app:regbose}

In this Appendix we comment on the regularization of the boson--boson
interaction and on how the bare coupling constant $\gBBtilde$ connects to the
scattering length $\aBB$ which parametrizes the s-wave scattering amplitude
$f(k)$ in the low-energy limit. Note that the following arguments apply equally
well to the boson--impurity coupling $\sim\gBFtilde$.  The relation between
$\gBBtilde$ and $\aBB$ is obtained by considering the two-body scattering
problem which in principle requires a rigorous treatment of the boson--boson
interaction by solving the Lippmann--Schwinger equation (LSE) using the actual
two-body interaction potential as an input. Solving this problem is in general
very difficult---in most cases not even the two-body potential is known
exactly---but fortunately it is unnecessary when one is only interested in
physics at low collision energies where the physics is described by the
scattering length $\aBB$ alone. Indeed, in the case of low-energy scattering the
solution of the LSE can be viewed as a mapping of the two-body potential on this
single parameter $\aBB$. Exploiting this consequence of universality, it is
possible to choose a representative out of the class of interaction potentials
leading to the same value of \aBB. A particularly convenient choice is the
contact interaction used in our model \eqref{eq:microscopic_action}.

Having properly set up the microscopic model~\eqref{eq:microscopic_action},
we are now interested in the connection between $\aBB$ and $\gBBtilde$. To this
end, it is sufficient to consider the scattering amplitude $f(k)$ at vanishing
scattering momentum $\vec k$. In this limit the evaluation of the LSE yields
(in this Appendix we use units where $\mB=1$) 
\begin{equation}
  \frac{1}{\Gamma_{\phi\phi}^\text{2B}(0,\vec 0)}=\frac{1}{4 \pi
  \aBB}=\frac{1}{\gBBtilde}+\int\frac{d^3q}{(2\pi)^3}  \frac{g(q)}{q^2}
  \label{eq:LS1} \end{equation}
where we used the relation between the T-matrix and the scattering amplitude
$f(k)=-\Gamma_{\phi\phi}^\text{2B}(\omega=\vec k^2,\vec 0)/4\pi$ for two atoms
scattering in the center-of-mass frame with momenta $\vec k$ and $-\vec k$. In
the case of contact interactions, $g(q)=1$ so that the second term in
Eq.~\eqref{eq:LS1} diverges and has to be regularized. The appearance of this
divergency is however a pure artifact which originates from the choice of the
rather unphysical contact interaction. Any typical interaction potential of a
finite range $r_0$ would automatically lead to a regularization of the integral
at a momentum scale $\Lambda$ of the order of the inverse range $1/r_0$. In
Eq.~\eqref{eq:LS1} this is mimicked by the presence of the function $g(q)$. The
regularization at high momenta can in particular be realized by introducing a
sharp momentum cutoff at the scale $\Lambda$ which for cold atoms is then of the
order of the inverse van-der-Waals length $1/l_\text{vdw}$. Carrying out the
integral in Eq.~\eqref{eq:LS1} one obtains 
\begin{equation}\label{eq:LS2}
  \frac{1}{\gBBtilde}=\frac{1}{4\pi\aBB}-c\Lambda\ , 
\end{equation} 
where  $c$ is a numerical factor which for the choice of a sharp momentum
regulator $g(q)=\theta(\Lambda-q)$ is given by $c=1/(2\pi^2)$. 

The simple Eq.~\eqref{eq:LS2} represents the connection between $\aBB$ and
$\gBBtilde$ we sought after and from which a few subtleties in the
renormalization procedure of the coupling $\gBBtilde$ become apparent. For
instance, assuming a microscopically repulsive interaction, i.e., $\gBBtilde>0$,
we directly see that in the zero-range limit $r_0\to 0 $, i.e.,
$\Lambda\to\infty$, the scattering length $\aBB$  is bound to be zero which
merely reflects the fact that any repulsive potential, no matter how strong, can
feature a scattering length at most as large as the range of the potential
\cite{landauqm}.  The only way to recover a finite scattering length for
repulsive interactions is thus to keep $\Lambda\sim 1/r_0$ finite.  From
Eq.~\eqref{eq:LS2} it is then apparent that if the Bose gas is very weakly
interacting with a small scattering length $a\Lambda\ll1$, one may neglect the
term proportional to $\Lambda$ which yields the commonly used relation
$\gBBtilde=4\pi\aBB$ we also employed in the Bogoliubov approximation of the
Bose gas discussed in Section \ref{sec:model}.

The above argument ignores the fact that cold atoms are always interacting via
microscopically \emph{attractive} interactions.  Thus, $\gBBtilde$ is bound to
be negative, no matter the sign of $\aBB$.  Although making the matter more
difficult, attractive interactions also lead to far richer physics. For
instance, by virtue of Eq.~\eqref{eq:LS2} it is possible to achieve any desired
scattering length $\aBB$ by fine-tuning $\gBBtilde$ as function of the range of
the potential $r_0\sim1/\Lambda$. Furthermore, for attractive interactions it is
possible to take the zero-range limit $\Lambda\to\infty$ while still retaining
an arbitrary value of $\aBB$. This also reflects the presence of universality:
by adjusting $\gBBtilde$ according to Eq.~\eqref{eq:LS2} and afterwards sending
$\Lambda$ to infinity, all information about the short-range details are hidden
in the single parameter $\gBB$.

\section{Analytical calculation of the T-matrix with finite boson--boson
interaction}
\label{app:tmatrix}

With some effort, it is possible to calculate an analytical expression for the
T-matrix even when including the Bogoliubov propagators of the weakly
interacting Bose gas instead of their bare counterparts. The calculation is most
conveniently done in units where masses are measured in units of the boson mass
$\mB$, energies in units of $\mu_\phi=\gBB\rhoBEC$, and consequently length
scales in units of the healing length $1/\sqrt{\gBB\rhoBEC}$.  In these units,
the pair propagator $L(\Omega,\vec p)$ only depends on $\Omega$ and $p=|\vec p|$
which simplifies the notation considerably. To restore the units used in the
main part of this article where we set the typical interparticle distance $\sim
n^{-1/3}$ to one instead of the healing length, one has to use the substitution
\begin{equation}
  L(\Omega,p)\mapsto \sqrt{\gBB}\,L\left(
  \frac{\Omega}{\gBB},\frac{p}{\sqrt{\gBB}} \right)\ .
  \label{eq:unit_substitution}
\end{equation}

The first step in the calculation of the pair
propagator~\eqref{eq:pair_propgator} is the evaluation of the frequency
integral. Since we only consider the limit of a vanishing impurity density, the
fermionic propagator's pole in imaginary frequency representation are in the
lower half plane. The frequency integration contour can thus be closed in the
upper
half plane so that only one of the two poles of the boson propagator \Gnormal
contributes.
After analytic continuation, the expression for the pair propagator reduces  to (in our new units)
\begin{multline}
  L(\Omega,\vec{p}) = -\int\frac{\d^3k}{(2\pi)^3}\left\{ 
  \frac{\sqrt{k^2(k^2+4)}+k^2+2}{2\sqrt{k^2(k^2+4)}}
  \right. \\ \left.
  \times\frac{1}{\Omega-\frac{1}{2}\sqrt{k^2(k^2+4)}
  -\frac{(\vec{p}-\vec{k})^2}{2\alpha}+i0^+}
  +\frac{2\alpha}{(\alpha+1)k^2}
  \right\}\ .
  \label{eq:fullL}
\end{multline}
In a second step, we separate $L$ into its real and imaginary part using the
identity $(x+i0^+)^{-1}=\mathcal{P}(1/x)-i\pi\delta(x)$, where $\mathcal{P}$
denotes the principal value. We carry out the angular part of the integration
which yields
\begin{multline}
  \Re L(\Omega,\vec{p}) = -\frac{1}{(2\pi)^2}
  \fint_0^\infty\d k \left\{\alpha 
  \frac{f(k)}{p}
  \log\left|
  \frac{g(\Omega,k,p)}{g(\Omega,k,-p)}
  \right|
  \right. \\ \left.
  +\frac{4\alpha}{\alpha+1}
  \right\}\ ,
  \label{eq:ReL_integral}
\end{multline}
where $\fint$ denotes the principal value integral, and
\begin{multline}
  \Im L(\Omega,\vec{p}) = \frac{\alpha}{4\pi}\int_0^\infty\d k \frac{f(k)}{p}
  \left\{ 
  \theta\left[ g(\Omega,k,p) \right]
  \right. \\ \left.
  -\theta\left[ g(\Omega,k,-p) \right]
  \right\}
  \label{eq:ImL_integral}
\end{multline}
with functions $f$ and $g$ defined by
\begin{equation}
  f(k) = k\frac{\sqrt{k^2(k^2+4)}+k^2+2}{2\sqrt{k^2(k^2+4)}}
  \label{eq:fk}
\end{equation}
and
\begin{equation}
  g(\Omega,k,p) = \Omega-\frac{1}{2}\sqrt{k^2(k^2+4)}-\frac{(p-k)^2}{2\alpha}\ .
  \label{eq:gkp}
\end{equation}

\subsection{Real part of the pair propagator}
\label{ssec:ReL}

Using the substitution $k=x-1/x$, Eq.~\eqref{eq:ReL_integral} becomes
\begin{multline}
  \Re L(\Omega,\vec{p}) = -\frac{1}{(2\pi)^2}\left\{ \frac{4\alpha}{\alpha+1}
  \right. \\ \left.
  + \fint_1^\infty\d x\,\left[ 
  \frac{\alpha x}{p}
  \log\left|\frac{h(\Omega,p,x)}{h(\Omega,-p,x)}\right|
  +\frac{4\alpha}{\alpha+1}
  \right]
  \right\}\ ,
  \label{eq:ReL_substituted}
\end{multline}
where 
\begin{multline}
  h(\Omega,p,x) = x^4-\frac{2p}{\alpha+1}x^3
  -\frac{2\alpha}{\alpha+1}\left(
  \Omega-\frac{p^2}{2\alpha}+\frac{1}{\alpha} \right)
  \\
  +\frac{2p}{\alpha+1}x - \frac{\alpha-1}{\alpha+1}
  \label{eq:p4}
\end{multline}
is a fourth-order polynomial in $x$.  It can easily be verified from the
expression for the function $h$ that its zeros $x_i$, $i=1,\dots,4$ (which may
be calculated analytically) obey the simple identity $\sum_ix_i=2p/(\alpha+1)$.
Due to the symmetry $h(\Omega,-p,x)=h(\Omega,p,-x)$, the zeros in the
denominator in Eq.~\eqref{eq:ReL_substituted} are obtained by multiplying $x_i$
by $-1$. 

In order to  progress further, we rewrite $h$ as
$h(\Omega,p,x)=\prod_i(x-x_i)$ and decompose the logarithm into a sum of
eight logarithms containing one factor each. The integrals to be solved are
then of the form $\fint_1^{\xmax}\d x\, x\log|x-x_i|$ which is analytically
tractable \footnote{Note the introduction of a finite upper limit $\xmax$.
Although the overall integrand is convergent, the terms in the sum are not
convergent individually. After integrating and summing up, the limit
$\xmax\rightarrow\infty$ may safely be taken.}. 
Writing $x_j=a_j+ib_j$ with $a_j,b_j\in \mathbb{R}$, carrying out the integral, rearranging the sum using the
identity $\sum_i x_i=2p/(\alpha+1)$ and taking the limit
$\xmax\rightarrow\infty$, one finally obtains the analytical result 
\begin{multline}
  \Re L(\Omega,\vec{p}) = -\frac{1}{(2\pi)^2} \Bigg\{ 
  \frac{2\alpha}{\alpha+1}+\frac{\alpha}{p}\sum_{j=1}^4\bigg[ 
  a_j|b_j|
  \\
  \times\left( \pi-\arctan\frac{1+a_j}{|b_j|}-\arctan\frac{1-a_j}{|b_j|}
  \right)
  \\ 
  +\frac{1-a_j^2+b_j^2}{2}\atanh\frac{2a_j}{1+a_j^2+b_j^2}
  \bigg]
  \Bigg\}\ .
  \label{eq:ReLreal}
\end{multline}

A considerable
simplification occurs for $\alpha=1$. Then the constant term in $h$
vanishes identically such that in the ratio $h(x)/h(-x)$ one
power of $x$ cancels, leaving us with the third-order polynomial
\begin{equation}
  h(\Omega,p,x) = x^3-px^2-\left(\Omega-\frac{p^2}{2}+1\right)x+p
  \label{eq:hx}
\end{equation}
with three complex zeros.  The result becomes even simpler for the case of $p=0$ where $h$
takes the form of a quadratic polynomial.  Carrying out the
calculation, one then finds the compact result
\begin{equation} 
  \Re L(\Omega)=\frac{-1}{2\pi^2} \begin{cases}
    1-\frac{\arctan\sqrt{-1-\Omega}}{\sqrt{-1-\Omega}}\Omega & \Omega < -1
    \\[.7em] 1-\frac{\atanh\sqrt{1+\Omega}}{\sqrt{1+\Omega}}\Omega &
    -1\leq\Omega\leq 0 \\[.7em]
    1-\frac{\atanh\frac{1}{\sqrt{1+\Omega}}}{\sqrt{1+\Omega}}\Omega & 0 < \Omega
  \end{cases}
  \label{eq:ReL_p0} 
\end{equation}
which we already quoted in a different set of units in Eq.~\eqref{eq:ReL_p0_rho_units}.

\subsection{Imaginary part of the pair propagator}
\label{ssec:ImL}

The calculation of the imaginary part of the pair propagator is
considerably simpler. Since the Heaviside functions inside the integrand in Eq.~\eqref{eq:ImL_integral} do
nothing but change the limits of integration, one is left with an integral over
$f(k)$ yielding
\begin{equation}
  \Im L(\Omega,\vec{p}) = \frac{\alpha}{16\pi p}\left(k^2+\sqrt{k^2(k^2+4)}\right)
  \bigg|_{k_{\text{min}}(\Omega,\vec{p})}^{k_{\text{max}}(\Omega,\vec{p})}\ ,
  \label{eq:ImL_explicit}
\end{equation}
where $k_{\text{min}}$ and $k_{\text{max}}$ denote the real and positive zeros of
the functions $g(\Omega,k,p)$ and $g(\Omega,k,-p)$ defined by 
Eq.~\eqref{eq:gkp}.

Studying the properties of the function $g$, one finds
the very simple result that for small external momenta the onset for a nonzero $\Im L$ as function of frequency for given momentum $p$ obeys the exact relation 
\begin{equation} 
  \Omegamin(p) = -g(0,0,p) =
  \frac{p^2}{2\alpha} \quad (p\leq\alpha)\ ,
  \label{eq:omega_min_small_p}
\end{equation}
which can be viewed as a consequence of the Landau criterion for the onset of superfluidity.

\begin{figure}[tbhp]
  \centering
    \includegraphics[width=\linewidth]{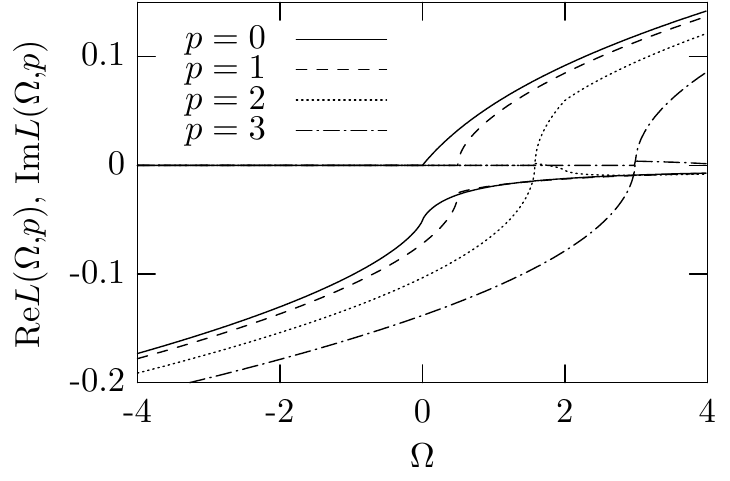}
  \caption{Real and imaginary part of the pair propagator $L$ for $\alpha=1$
  and different values of the external momentum $p$. $\Im L$ is strictly
  positive while $\Re L$ is mostly negative. Note that this figure uses units in
  which $\gBB\rhoBEC=1$.}
  \label{fig:L}
\end{figure}
As in the case of the real part, it is insightful to consider the limiting case
$p=0$. In this limit, $k_{\text{min}}$ and $k_{\text{max}}$ converge towards the
same value and the expression in Eq.~\eqref{eq:ImL_integral} becomes a
derivative with respect to $p$. Calculating the derivative and then carrying out
the integral one finds
\begin{equation}
  \Im L(\Omega,0)=\frac{k_0}{4\pi}
  \frac{\sqrt{k_0^2(k_0^2+4)}+k_0^2+2
     }{
     \alpha^{-1}\sqrt{k_0^2(k_0^2+4)}+k_0^2+2}\ ,
  \label{eq:ImL_p0}
\end{equation}
where $k_0=k_0(\Omega)$ is the solution to the equation $g(\Omega,k_0,0)=0$
which is readily calculated analytically. Again, the expression becomes
particularly simple in the case $\alpha=1$ where one finds [cf.~Eq.~\eqref{eq:ImL_alpha_1_p0_rho_units}]
\begin{equation}
  \Im L(\Omega,0) = \frac{1}{4\pi}\frac{\Omega}{\sqrt{1+\Omega}}\theta(\Omega)
  \ .
  \label{eq:ImL_alpha_1_p0}
\end{equation}

To illustrate the results of this Appendix, we show the real and imaginary part of the pair propagator for the mass-balanced case $\alpha=1$ in Fig.~\ref{fig:L}.

\end{document}